\documentclass[graybox]{svmult}

\usepackage{mathptmx}       
\usepackage{helvet}         
\usepackage{courier}        
\usepackage{type1cm}        
\usepackage{graphicx}        
\usepackage{pstricks}
\usepackage{multicol}        
\usepackage[bottom]{footmisc}

\makeindex             

\newcommand{\vc}[1]{\textbf{\em #1}}

\newcommand{\apjl}{Astrophys. J. Lett.}
\newcommand{\araa}{Annu. Rev. Astron. Astrophys.}

\newcommand{\grl}{Geophys. Res. Lett.}
\newcommand{\jgr}{J. Geophys. Res.}
\newcommand{\jpcs}{J. Phys. Conf. Ser.}
\newcommand{\mnras}{Mon. Not. R. Astron. Soc.}
\newcommand{\ssr}{Space Sc. Rev.}
\newcommand{\pop}{Phys. Plas.}

\begin{document}

\title*{Particle Acceleration by Magnetic Reconnection}

\author{Elisabete M. de Gouveia Dal Pino and Grzegorz Kowal}
\institute{E. M. de Gouveia Dal Pino \at
Instituto de Astronomia, Geof\'\i sica e Ci\^{e}ncias Atmosf\'ericas,
  Universidade de S\~ao Paulo, \\ Rua do Mat\~ao, 1226 -- Cidade Universit\'{a}ria, \\
 05508-090, S\~ao Paulo/SP, Brazil \\
\email{dalpino@iag.usp.br}
\and  G. Kowal \at
Escola de Artes, Ci\^{e}ncias e Humanidades, Universidade de S\~ao Paulo, 
\\ Av. Arlindo B\'ettio, 1000 -- Ermelino Matarazzo, 
03828-000, S\~ao Paulo/SP, Brazil \\
\email{grzegorz.kowal@usp.br}
}

%
%

\maketitle

\abstract*{
Observational data require a rich variety of mechanisms to accelerate fast particles in astrophysical environments operating under different conditions.  The mechanisms discussed in the literature include varying magnetic fields in compact sources, stochastic processes in turbulent environments, and acceleration behind shocks. An alternative, much less explored mechanism  involves particle acceleration within magnetic reconnection sites. In this chapter we discuss this mechanism and show that particles can be efficiently accelerated by reconnection through a first order Fermi process within large scale current sheets (specially when in the presence of local turbulence which speeds up the reconnection and  make  the acceleration region thicker)  and also through a second order Fermi process in pure MHD turbulent environments.
}

\abstract
{Observational data require a rich variety of mechanisms to accelerate fast particles in astrophysical environments operating under different conditions.  The mechanisms discussed in the literature include varying magnetic fields in compact sources, stochastic processes in turbulent environments, and acceleration behind shocks. An alternative, much less explored mechanism  involves particle acceleration within magnetic reconnection sites. In this chapter we discuss this mechanism and show that particles can be efficiently accelerated by magnetic reconnection through a first order Fermi process within large scale current sheets (specially when in the presence of local turbulence which speeds up the reconnection and  make  the acceleration region thicker)  and also through a second order Fermi process in pure MHD turbulent environments.
}

\section{Introduction}
\label{sec:Intro}

Energetic particles are ubiquitous in astrophysical environments and their
acceleration still challenges the researchers. For instance, the origin of the
ultra high energy cosmic rays (UHECRs) is  unknown yet.  Their spectrum is
consistent with an origin in extragalactic astrophysical sources and candidates
range from the birth of compact objects to explosions related to gamma-ray
bursts (GRBs), or events in active galaxies (AGNs) \cite{kotera11}, however, the mechanism(s) that produce(s) them is(are) still not fully understood.
Similarly, recent very high
energy observations  with the Fermi and Swift satellites and
ground based gamma ray observatories (HESS, VERITAS and MAGIC) of AGNs and GRBs have been challenging the current
 particle acceleration theories
which have to explain how particles are accelerated to TeV or larger energies
in regions relatively small compared to the fiducial scale of their sources
\cite{sol11}.

The mechanisms frequently discussed in the literature for accelerating energetic particles include varying magnetic fields in compact sources (e.g., \cite{degouveia00, degouveia01, melrose09}), stochastic processes in turbulent environments \cite{melrose09}, and acceleration behind shocks.
  The latter, in particular, has been extensively discussed in the literature \cite{sironi09, melrose09, kotera11}
An alternative, much less explored mechanism so far, involves particle acceleration within magnetic reconnection sites.

Magnetic reconnection occurs when two magnetic fluxes of opposite polarity encounter each other (see middle panel of Figure 7). In the presence of finite magnetic resistivity, the converging magnetic field lines annihilate at the discontinuity surface and a current sheet forms there.

Traditionally, particle acceleration in reconnection sites has been regarded as a linear process 
due to the advective electric field (also referred as the reconnection electric field) that develops along the current sheet, in the normal direction to the magnetic field ($\epsilon = V_R B/c$, where $V_R$ is the reconnection velocity) (e.g., \cite{speiser65, litvinenko96, zenitani01, giannios10}).  While describing  a betatron-like orbit (also often called
Speiser orbit) along this direction, particles are continuously 
accelerated by the electric field  with its energy increasing linearly
with the distance (z)  travelled along the current sheet or reconnection layer ($E \sim  e V_R B z/c$) \cite{speiser65}.

In 2005, de Gouveia Dal Pino \& Lazarian \cite{degouveia05}(henceforth GL05)   proposed a mechanism to accelerate particles to relativistic velocities within the reconnection layer, in a similar way to the first-order Fermi process that occurs in shocks, which is also able to increase their energy exponentially.
It is known from shock acceleration theory that particles are injected upstream and
allowed to convect into the shock, while diffusing in space so as to undergo
multiple shock crossings, and thereby gaining energy through a
 first order Fermi
process \cite{fermi49}.  Similarly, GL05 \cite{degouveia05}
 proposed that trapped  charged particles may bounce back and forth several times and gain energy due to head-on collisions with the two converging magnetic fluxes of opposite polarity that move to each other at the reconnection velocity ($V_{R}$).
They found that the particle
energy gain after each round trip is $\Delta E/E \propto V_{R}/c$.  Under {\em
fast} magnetic reconnection conditions, e.g. induced by turbulence
\cite{lazarian99}, $V_{R}$ can be of the order of the local Alfv\'en speed
$V_{A}$ (see below and also the Chapter by Lazarian et al. in this volume).
  At  the surroundings of relativistic sources, for instance, $V_{R} \simeq  v_A \simeq c$, so that the mechanism can be rather efficient. GL05 \cite{degouveia05} have also shown that the accelerated particles have a power-law distribution  and a corresponding electron synchrotron radio power-law spectrum  which is compatible with the observed radio flares of galactic black hole binaries (microquasars). Though that study was specifically applied to microquasars, it can be far more general in astrophysical systems, as we discuss below.

   Afterwards, Drake et al. (2006) \cite{drake06}  invoked a similar process, but within a collisionless reconnection scenario (see the chapter by Lazarian et al. in this volume).  In their model, the contraction of two-dimensional magnetic loops is controlled by
  the firehose instability that arises in a particle-in-cell (PIC) domain
  \cite{lyubarsky08,drake10}.  Other processes of acceleration, e.g. due to
turbulence arising as a result of reconnection
\cite{larosa06} were shown to be less dominant.

Magnetic reconnection is very frequent and therefore, it should be expected to induce acceleration of particles in a wide
range of galactic and extragalactic environments.  Originally discussed
predominantly in the context of electrons in solar flares \cite{drake06,
drake09, gordovskyy10, gordovskyy11, zharkova11}, it was later applied to
explain the origin of anomalous cosmic ray protons \cite{lazarian09,
drake10}, and the anisotropies in the direction of solar system
magnetotail \cite{lazarian10}.  It also has been gaining importance beyond the solar system, in more
extreme astrophysical environments and sources, such as in the production of  ultra
high energy cosmic rays  \cite{giannios10,
kotera11}, in particle
acceleration in jet-accretion disk systems \cite{degouveia05, degouveia10a, degouveia10b,
giannios10, delvalle11}, and in the general framework of compact sources, as AGNs and GRBs
\cite{lazarian03, zenitani01, zenitani09, degouveia10b, giannios10, zhang11, uzdensky11a, uzdensky11b,
degouveia11}, and even in pulsar nebulae, like Crab \cite{cerutti13}.

The applications above, however, still require extensive study of
 particle acceleration in magnetic reconnection sites, as well as on  its
connection with magnetohydrodynamical (MHD) turbulence and $fast$ magnetic
reconnection.

In particular, a way to probe the  analytical results above is through numerical simulations. So far, most of the numerical studies of particle acceleration by magnetic reconnection have been  
performed  for two-dimensional, collisionless pair plasmas by means of particle-in-cell (PIC) simulations    (e.g., \cite{drake06, drake12, zenitani01, cerutti13}). However, these apply to kinetic scales of only a few hundred  plasma inertial lengths  ($\sim 100 c/\omega_p$, where $\omega_p$ is the plasma frequency). The generally much 
larger scales of the astrophysical systems (pulsars, AGNs, GRBs, etc) frequently require a collisional MHD description of reconnection.


Some progress  in
this direction has been achieved recently  \cite{degouveia10b, degouveia11, lazarian11, kowal11, kowal12}   where the  model of GL05
\cite{degouveia05} was  tested successfully by means of two (2D) and three dimensional (3D) MHD simulations.  Kowal, de Gouveia Dal Pino \& Lazarian (2011, henceforth LGK11) \cite{kowal11}, in particular, have  shown that the
acceleration of particles inserted in MHD domains of reconnection without
including kinetic effects produces results similar to those found in particle-in-cell (PIC)
(collisionless) simulations where particle acceleration is controlled by kinetic
effects such as the firehose instability \cite{zenitani01, drake06, drake10}.  This demonstrated  that the
acceleration in reconnection regions is a universal process which is not
determined by the details of the plasma physics and can be also very efficient
in collisional gas.  They have also shown that particle acceleration in 3D
MHD reconnection behaves quite differently from the acceleration in 2D domains
since the increase in the acceleration component parallel to the magnetic
field is not constrained by the production or size of contracting islands, as in
the 2D case.  These results call for focusing on realistic 3D geometries of
reconnection.

Other concomitant studies have also explored test particle acceleration in MHD domains \cite{gordovskyy10,gordovskyy11}.  Gordovskyy et al. (2010) \cite{gordovskyy10}, for instance, focussed on 2D models with  time-dependent reconnection, while Gordovskyy and  Browning (2011) \cite{gordovskyy11}, aiming  the study of particle acceleration in small solar flares,  examined  a somewhat different scenario, with the acceleration of test particles by magnetic reconnection induced by kink instabilities in 3D twisted magnetic loops. Although they have obtained results for the particle energy  distributions which are compatible with field-aligned acceleration as in the studies above, they did not explore the nature of the mechanism  accelerating  the particles.

Kowal, de Gouveia Dal Pino \& Lazarian (2012, henceforth KGL12)\cite{kowal12} have, in turn, injected
 test particles in different 3D collisional MHD reconnection and
compared the particle spectrum and acceleration rates in these different domains.
When
considering a single Sweet-Parker topology \cite{sweet58,parker57} (subject to large artificial magnetic resistivity to allow fast reconnection), they have found that particles accelerate
predominantly through a first-order Fermi process, as predicted in GL05  \cite{degouveia05}.  When turbulence is induced within the
current sheet, the acceleration  is
highly enhanced.  This is due to the fact that reconnection becomes fast in a natural way (and
independent of magnetic resistivity) in the presence of turbulence
\cite{lazarian99, kowal09}  and allows the formation of a thick volume filled
with multiple simultaneously reconnecting magnetic fluxes. 
Besides, reconnection is intrinsically 3D in this case (see also the chapter by Lazarian et al. in this volume for more details of the fast reconnection process in the presence of turbulence).
The particles trapped
within this volume suffer several head-on scatterings with the contracting
magnetic fluctuations, which significantly increase the acceleration rate  and the amount of particles
 which are accelerated through   a first-order Fermi process.  They have  also tested the
acceleration of particles in  pure MHD turbulence, where particles suffer
collisions both with approaching and receding magnetic irregularities. The
acceleration rate is smaller in this case and suggests that the dominant process is a
second order Fermi.

In this chapter, we discuss these acceleration mechanisms in magnetic reconnection sites  in detail and review the recent analytical and numerical results in this regard.

\section{Analytical model for first order Fermi particle acceleration within magnetic reconnection sites}
\label{sec:analytic}

\begin{figure}[t]
\begin{center}
\includegraphics[scale=.5]{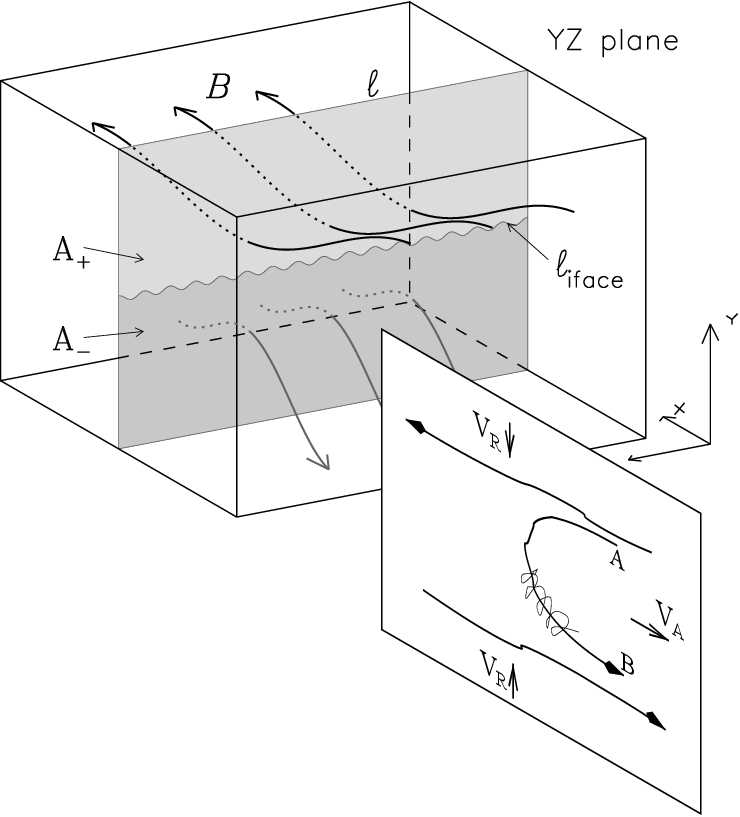}
\caption{
Particle acceleration in  a reconnection site where two magnetic field fluxes of opposite polarity move to each other. 
Top left: three-dimensional view of a magnetic reconnection sheet. Bottom right: detail of a particle being accelerated within the reconnection site. It spirals about a reconnected magnetic
field line and bounces back and forth  between points A and B. The reconnected
regions move towards each other with the reconnection velocity
$V_R$. Particles gain energy due to "collisions" with the magnetic irregularities within the two converging fluxes, just like in the first-order Fermi process in shock fronts (see GL05 \cite{degouveia05}). 
Bouncing between the points A and B happens
 either because of streaming instability induced by energetic particles or by magnetic
turbulence in the
reconnection region (as discussed in \cite{lazarian99}). In particular, when turbulence is present, the acceleration region is filled in by several  oppositely moving reconnected flux tubes  which collide and  repeat on  smaller and smaller scales the pattern of the larger scale reconnection making the process very fast and therefore, the particle acceleration very efficient. Since in such a case the reconnection is naturally a three-dimensional process, particle acceleration as shown in the bottom right panel may occur in all directions within the current sheet and is not restricted to the direction  depicted in the figure.   (Adapted from  \cite{kowal09, lazarian11}; see also \cite{degouveia14}). 
}
\label{fig_recon}
\end{center}
\end{figure}

We first  discuss  an analytical model for   acceleration of particles within reconnection sites which was originally introduced by  GL05 \cite{degouveia05}  (see also \cite{lazarian05, degouveia10a, degouveia14}. Figure~\ref{fig_recon} illustrates  the simplest realization of the acceleration within a large scale reconnection region. As described in the Figure, as a particle bounces back and forth between two converging magnetic fluxes of opposite polarity, it gains energy through a first-order Fermi acceleration.

In order to derive the energy spectrum of the accelerated particles one can invoke a similar procedure to the one employed in the calculation of the first order Fermi acceleration in shocks (see, e.g., \cite{bell78, longair92, lazarian11}). Let us consider the  acceleration of $M_0$ particles with an initial energy $E_0$. If a particle acquires an energy $E=\beta E_0$ after a collision, its energy after $m$ collisions will $\beta^m E_0$. At the same time if the probability of a particle to remain within the acceleration region is $P$, after $m$ collisions the number of accelerated particles will be $M = P^m M_0$. Therefore,
 $\ln (M/M_0)/\ln(E/E_0)=\ln P/\ln\beta$ or
\begin{equation}
\frac{M}{M_0}=\left(\frac{E}{E_0}\right)^{\ln P/\ln\beta}
\end{equation}
Since some of these $M$ particles will be further accelerated before escaping the system, the equation above implies that the number $N(E)$ of particles accelerated to energies equal to or larger than $E$ is given by:
\begin{equation}
N(E)dE=const\times E^{-1+(\ln P/\ln\beta)} dE
\label{NE}
\end{equation}

To compute $P$ and $\beta$ within  the reconnection site we  may consider the following process. The particles from the upper reconnection region will "see" the lower reconnection region moving towards them with the velocity $2V_{R}$ (see Figure~\ref{fig_recon}). If a particle from the upper region enters at an angle $\theta$ with respect to the direction of $V_R$ into the lower region then the expected energy gain of the particle is $\delta E/E=2V_{R}\cos\theta/c$. For an isotropic distribution of particles their probability function is $p(\theta)=2\sin\theta\cos\theta d\theta$ and therefore the average energy gain per crossing of the reconnection region is
\begin{equation}
\langle \delta E/E \rangle =\frac{V_{R}}{c}\int^{\pi/2}_{0} 2\cos^2\theta \sin\theta d\theta=4/3\frac{V_{R}}{c}
\end{equation}
Particles will complete a  full  acceleration cycle when they return back to the upper reconnection region. Similarly, if they are  in the lower reconnection region they will see the upper reconnection region moving towards them with the speed $2 V_{R}$. As a result, a full acceleration cycle provides an energy increase $\langle \delta E/E \rangle_{cycle}=8/3(V_{R}/c)$ and thus
\footnote{We note that Giannios (2010) \cite{giannios10} re-derived the relation above in the limit when the reconnection velocity itself approaches the light speed and obtained an expression that naturally recovers the form of eq. 4 in the non-relativistic regime. }
\begin{equation}
\beta=E/E_0=1+8/3(V_{R}/c)
\label{beta}
\end{equation}

Let us assume that the particle diffusion velocity is much smaller than $V_R$. In analogy to particle acceleration in a  shock front, for simplicity, we further assume  that the total number of particles crossing the boundaries of the upper and lower magnetic fluxes is $2\times 1/4 (n c)$, where $n$ is the number density of particles. If the particles are advected to outside of the reconnection region with the magnetized plasma outflow then, the loss of the energetic particles will be given approximately by $2 V_{R}n$. Therefore the fraction of energetic particles which are lost in a cycle will be  $V_{R} n/[1/4(nc)]=4V_{R}/c$ and
\begin{equation}
P=1-4V_{R}/c.
\label{P}
\end{equation}

Combining Eqs.~(\ref{NE}), (\ref{beta}), and (\ref{P}) one obtains
\begin{equation}
N(E)dE=const_1 E^{-5/2}dE,
\label{-5/2}
\end{equation}
which is the spectrum of accelerated energetic particles for the case when the plasma back-reaction is negligible  \cite{degouveia05, lazarian11}.
We note  that the power-law index obtained above is independent of the reconnection velocity $V_R$. This is in part  due to the simplified assumptions in the derivation above. Nonetheless, we will see below that in the much more realistic numerical simulations of the acceleration of test particles in non-relativistic MHD reconnection sites, the acceleration rate and power spectrum are not very much sensitive to the reconnection speed \cite{delvalle13}.

In recent work, Drury (2012) \cite{drury12} tried to improve the analytical model above by considering two additional effects. First,    he took into account
the energy losses due to the outflow from the reconnection region
(which is associated with a divergence of
the flow field) and second, he  relaxed the assumption considered above (GL05) that the escape rate is the same as that from a
shock. Then, he repeated the calculation above and obtained a power law spectral index which is
 the same as in shock acceleration if  expressed  in terms of the compression ratio in the
system $r = \rho_2/\rho_1$, where $\rho_1$ and $\rho_2$ are the plasma densities at the inflow and the outflow regions of the reconnection site, respectively. In other words, he obtained:

\begin{equation}
f(p) \propto  p^{\frac{-3r}{r-1}},
\end{equation}
where $f(p)$ is the particles distribution function as a function of their momentum $p$. Considering that only
 little energy is used to heat the
plasma and the conversion is essentially one of magnetic energy into kinetic energy, then  the reconnection can be very compressive, unless the outflow is significantly  over pressured relative to the environment. But, for
a strongly magnetized inflow this is a very weak constraint \cite{drury12}. Thus one can expect  in general   large values of
the compression ratio, possibly larger than the value four usually assumed for adiabatic shocks. For  large values of $r$  one finds $f(p) \propto  p^{-3}$, or:

\begin{equation}
N(E) \propto E^{-1}
\end{equation}
which is a power-law spectrum much harder than the one obtained above by GL05, but confirms their prediction that  a rather efficient first-order Fermi particle acceleration process can take place in magnetic reconnection sites.

The considerations above also allow one to estimate
 the acceleration
time-scale due to reconnection, which is a  straightforward  generalization  of  the result  for shock
acceleration as well.

The simplest way to evaluate the acceleration time is 
by setting the energy of the accelerated particle $E$ equal to $e (V_{R}/c)  B z$, where
z  is the distance travelled by the particle along the current sheet (normal to the magnetic field direction) while being accelerated by the effective electric field $(V_{R}/c)  B$. The acceleration time is, therefore (e.g. \cite{speiser65, giannios10})

\begin{equation}
t_{acc} \simeq  z/c \simeq \frac{E } {e V_{R} B}.
\end{equation}



This acceleration time scale is similar to that for shock
acceleration, and the constraints on maximum energy due
to  the  finite  age  and  size  of  the  reconnection  region  will
thus be comparable to those in shock acceleration.

A simple way to estimate the maximum energy that a particle can achieve  is by realizing that it  can no longer be confined within the reconnection region when its Larmor radius becomes larger than the thickness of the reconnection layer $l_{rec}$. This  implies that:
\begin{equation}
E_{max} \simeq  e c l_{rec}  B 
\end{equation}

It should be  noticed that Drury's model above \cite{drury12} predicts that the acceleration within reconnection sites requires, as in shocks, a large compression ratio in order to be efficient. However, according to the discussion in the previous section  (see also  the following sections), the requirement for the magnetic reconnection acceleration process to proceed efficiently is to keep the accelerated particles within  contracting magnetic loops. This requires constraints on the particle diffusivity specially  perpendicular to the magnetic field direction.   The subtlety of this point is related to the fact that while in the first-order Fermi acceleration in shocks  compression is important, the acceleration via  reconnection  is  applicable also to incompressible fluids. Thus, unlike shocks, not the entire volume has to shrink by compression in order to the acceleration to occur, but  the volume of the magnetic flux tube. If the  perpendicular diffusion of the particles to the magnetic field is large they may decouple from the
magnetic field. Indeed, it is easy to see that as long as  the particles in the magnetic flux rope, as  depicted in Figure~1, bounce back and forth between the converging mirrors they will be accelerated. However,  if these particles leave the flux rope
too fast, they may start bouncing  between the magnetic fields of different flux ropes which may sometimes decrease their energy, thus favouring a second order rather than a first-order Fermi process. Thus it is important that the particle diffusion parallel and perpendicular to the magnetic field stays different. Particle anisotropy which arises from particle preferentially being accelerated in the parallel direction must be significant. In the next section where we depict results of numerical simulations of particle acceleration by magnetic reconnection, the evolution of both  particles' velocity components is tracked separately in order to stress this point.

In the case of laminar reconnection sites (i.e., with no turbulence), as in a Sweet-Parker model for reconnection (or as in the model above;  see also Section 3.4), the  scales  can be much smaller compared
to shock structures. This  will  lead to lower maximum energies (the dominant limiting loss process being diffusion out of the sides of the reconnection region).
This means that the models  developed above  apply only if  the particle diffusion
length scale in the inflow is
small compared to its lateral
extent. Thus they may only apply in a restricted energy
range which  should be contrasted with the case of shock acceleration where the scale separation is, in general, much larger. Nevertheless,  when turbulence is present within the reconnection site, this will make the reconnection volume much larger \cite{lazarian99, kowal09, kowal11, kowal12} and therefore,  this process can be competitive to shock acceleration, as we will see below.

\section{Particle Acceleration in reconnection sites: numerical studies }
\label{sec:numerics}

As discussed in the previous section,
magnetic reconnection results in shrinking of magnetic loops between two converging magnetic fluxes of opposite polarity and the charged particles
entrained over the magnetic loops are accelerated (see Figure 1).
In this section we discuss the results of numerical studies that confirm these predictions.

In what follows, we consider different domains of
magnetic reconnection  which were  modelled by solving the isothermal MHD equations
numerically in a uniform mesh using a
 Godunov-type scheme \cite{kowal09, kowal11, kowal12}.

In order to integrate the test particle trajectories  a data
cube obtained from the MHD models is frozen in time and then  10,000 test particles are injected in the domain
with random initial positions and directions and with an initial thermal
distribution.  For each particle  the relativistic  equation of motion is solved
\begin{equation}
 \frac{d}{d t} \left( \gamma m \vc{u} \right) = q \left( \vc{E} + \vc{u} \times
 \vc{B} \right) ,
 \label{eq:ptrajectory}
\end{equation}
where $m$, $q$ and $\vc{u}$ are the particle mass, electric charge and velocity,
respectively, $\vc{E}$ and $\vc{B}$ are the electric and magnetic fields,
respectively, $\gamma \equiv \left( 1 - u^2 / c^2 \right)^{-1}$ is the Lorentz
factor, and $c$ is the speed of light.  The electric field $\vc{E}$ is taken
from the MHD simulations
\begin{equation}
 \vc{E} = - \vc{v} \times \vc{B} + \eta \vc{J} , \label{eq:efield}
\end{equation}
where $\vc{v}$ is the plasma velocity, $\vc{J} \equiv \nabla \times \vc{B}$ is
the current density, and $\eta$ is the Ohmic resistivity coefficient.  The  resistive term above can be neglected because its effect on particle acceleration is
negligible \cite{kowal11}.  These  studies do not include the
particle energy losses, so that particles  gain or loose energy only through the
interactions with the moving magnetized plasma.

We note that  since
we are focusing on the acceleration process only,  very simple
domains can be considered which represent only small periodic boxes of entire magnetic
reconnection or turbulent sites.  For this reason, the typical crossing time
through the box of an injected thermal particle is very small and it has to
re-enter the computational domain several times before gaining significant
energy by multiple scatterings.  Thus, whenever a particle reaches the box
boundary it re-enters in the other side to continue scattering \cite{kowal12}.

\subsection{Particle Acceleration in 2D domains}

\begin{figure*}[ht]
 \center
 \includegraphics[width=\textwidth]{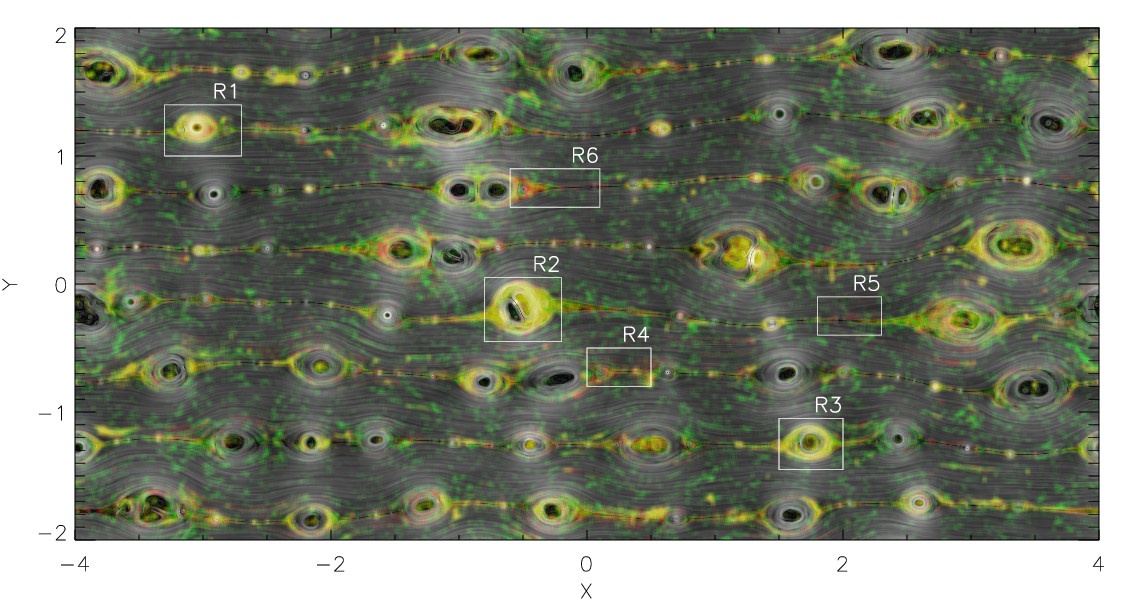}
 \caption{Topology of the magnetic field represented as a gray texture with
semi-transparent color maps representing locations where the parallel and
perpendicular particle velocity components are accelerated for a 2D model with
$B_z = 0.0$ at time $6.0$ in the code units.  The red and green colors
correspond to regions where either parallel or perpendicular acceleration
occurs, respectively, while the yellow color shows locations where both types of
acceleration occur.  The parallel component increases in the contracting islands
and in the current sheets as well, while the perpendicular component increases
mostly in the regions between current sheets.  White boxes show regions that are
more carefully analyzed in \cite{kowal11} paper.  The simulation was performed with the
resolution 8192x4096.   10,000 test particles were injected in this snapshot with
the initial thermal distribution with a temperature corresponding to the sound
speed of the MHD model. (From \cite{kowal11}.) }
 \label{fig:locations}
\end{figure*}

Figure~\ref{fig:locations} presents an evolved 2D MHD configuration with  eight Harris
current sheets in a periodic box \cite{kowal11} (see also \cite{drake10}).  The initial density
profile is  such that the total (gas plus magnetic) pressure is
uniform. Random weak velocity fluctuations were imposed to this
environment in order to enable spontaneous reconnection events and the
development of the magnetic islands.

\begin{figure*}[ht]
 \center
 \includegraphics[width=0.40\textwidth]{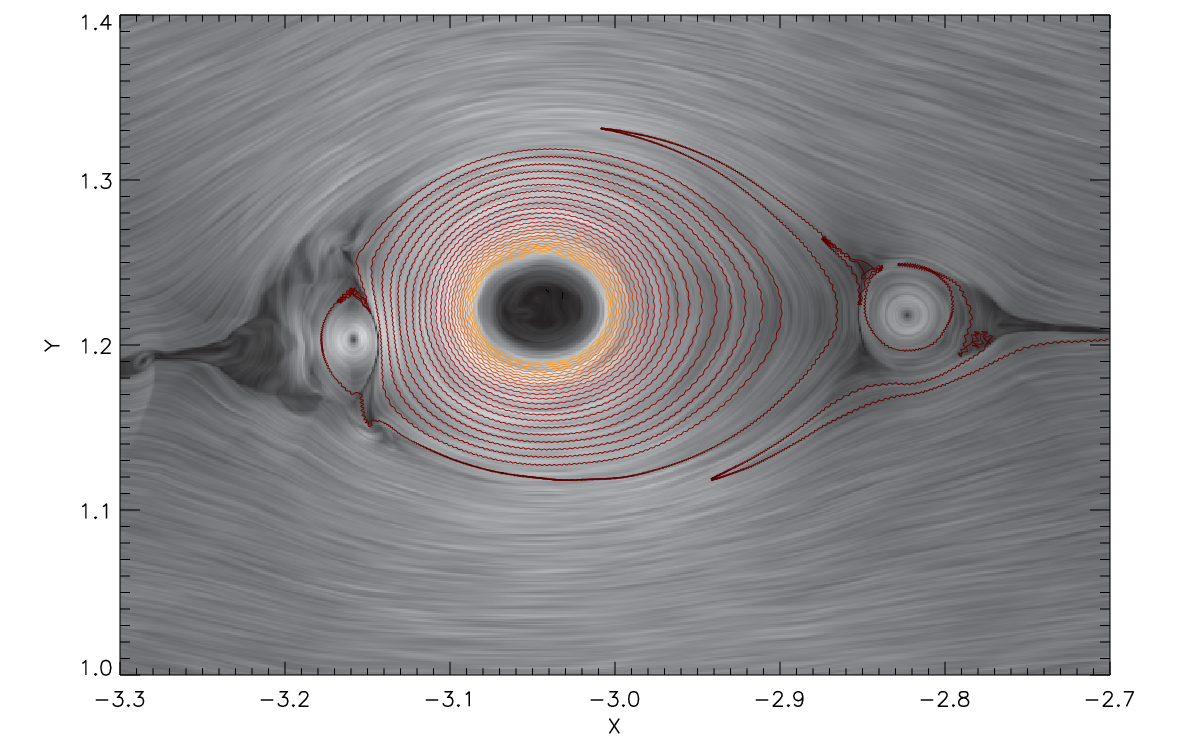}
 \includegraphics[width=0.30\textwidth]{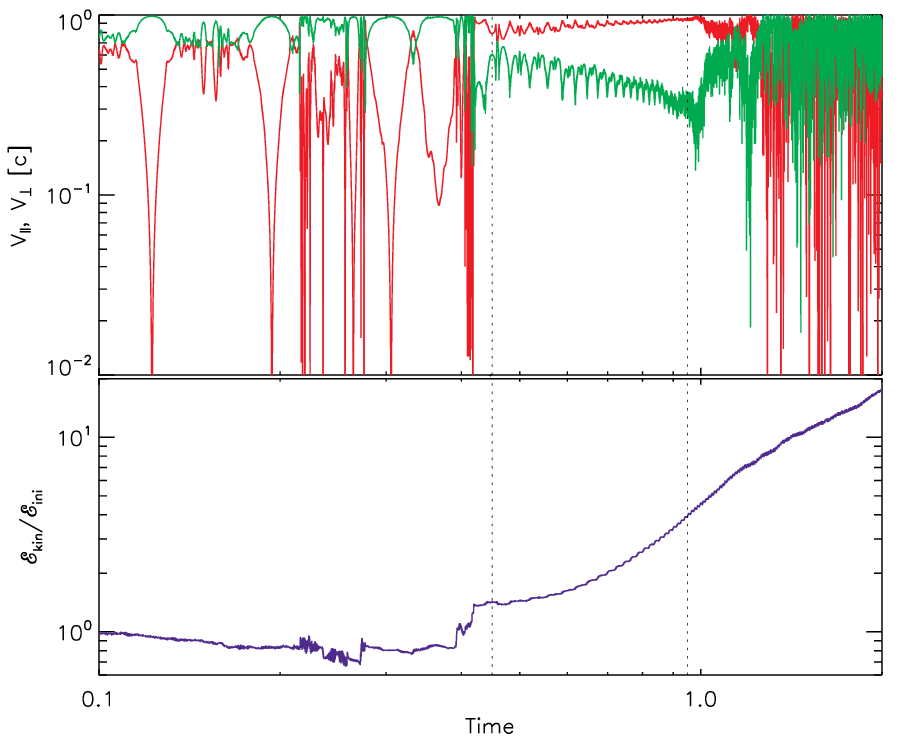}
 \caption{The case of a shrinking island where the particles accelerate
efficiently. It corresponds to region R1 of Fig.~\ref{fig:locations}.  The left panel shows
the trajectory of a single test proton trapped in this contracting island.  We see two
small magnetic islands on both sides of the central elongated island which are
merging with it.  This process results in the contraction of the central island.
The right panel  shows the exponential increase
of the particle energy.    The proton
orbiting around the center of the magnetic island increases its energy increment
after each orbit. (Extracted from \cite{kowal11}.) }
\label{fig:event}
\end{figure*}

Figure~\ref{fig:locations} clearly shows the merging of islands in
some locations and the resulting stretching  or shrinking  which provides
appropriate conditions for particle acceleration.
KGL11 \cite{kowal11} find
 that an increase of the parallel velocity component is mostly observed
within shrinking  islands and in current sheets (see the red and yellow zones in
Figure~\ref{fig:locations}), while the increase of the perpendicular component
is observed mostly near and within stretching  islands and between current sheets (see
the green and yellow zones in Figure~\ref{fig:locations}).  This complex
behavior is related to the degree of island deformation and the particle
direction and speed.
Within contracting magnetic islands or
current sheets the particles accelerate predominantly through the first order
Fermi process, as previously described, while outside of the current sheets and
the islands the particles experience mostly drift acceleration due to magnetic
fields gradients \cite{kowal11}.
In Figure~\ref{fig:event} the first of these effects is   zoomed in  an example of a single test proton which is trapped
in a shrinking  island and is accelerated.     Its parallel speed
increases while the gyro rotation slows down.  This results in an exponential
growth of the kinetic energy of the particle (as shown in the right panel).

Similar results were found in 2D collisionless pair plasma PIC simulations \cite{drake10, drake12, zenitani01, cerutti13}. 
In such cases, reconnection is  fast because it is facilitated by the two-fluid (Hall) effects and/or anomalous resistivity and exhibit a Petschek-like (1964) \cite{petschek64} structure
(\cite{birn01, yamada10, shay98, shay04, yamada06}).

This implies that the
first-order Fermi acceleration process within shrinking islands is not restricted to  collisionless physics or kinetic effects as previously suggested  and  described by PIC simulations (e.g. \cite{drake06, drake10, drake12} and references therein).
This acceleration mechanism in reconnection sites works also in collisional plasmas, under the MHD approximation, as shown above and, in fact,
MHD codes present an easier way to study the physics of particle acceleration
numerically.

\subsection{Acceleration Near and Within  Current Sheets}
\label{sec:current_sheets}

In the current sheet zones (regions R4 to R6 of Figure~\ref{fig:locations}) we can also identify
a first order Fermi acceleration due to
simple particle scattering between the converging flows entering both sides of
the current sheet (or even in merging/shrinking islands which are just forming there), as described in  \cite{degouveia05}.

In zones above
and below the current sheets particles possibly experience predominantly a drift
acceleration driven by non-uniformities of the magnetic field (e.g., \cite{melrose09}).  Generally, this
effect is less efficient than the first order Fermi process in
merging/contracting islands and results in smaller acceleration rates.  The origin of this effect is
due to the net work done on a charge by the Lorentz force (Eq.~\ref{eq:efield})
in a zone of non-uniform large scale magnetic field.  The principal equation
governing this is the scalar product of the particle velocity (or momentum) and
the acceleration by the convective electric field, $- \vc{v} \times  \vc{B}$.
In uniform magnetic fields, the energy gain and loss acquired during a
gyroperiod exactly cancel, so in result no net work is done, $\Delta W = 0$.

\begin{figure*}[ht]
 \center
 \includegraphics[width=0.40\textwidth]{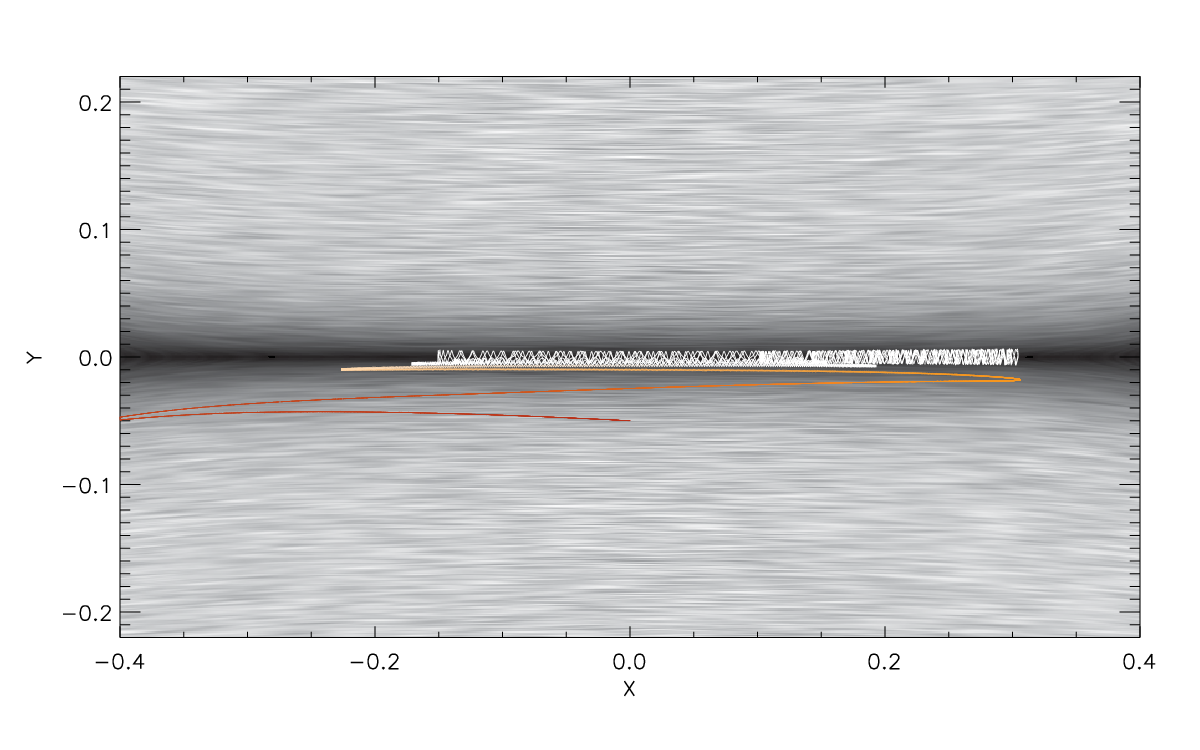}
 \includegraphics[width=0.30\textwidth]{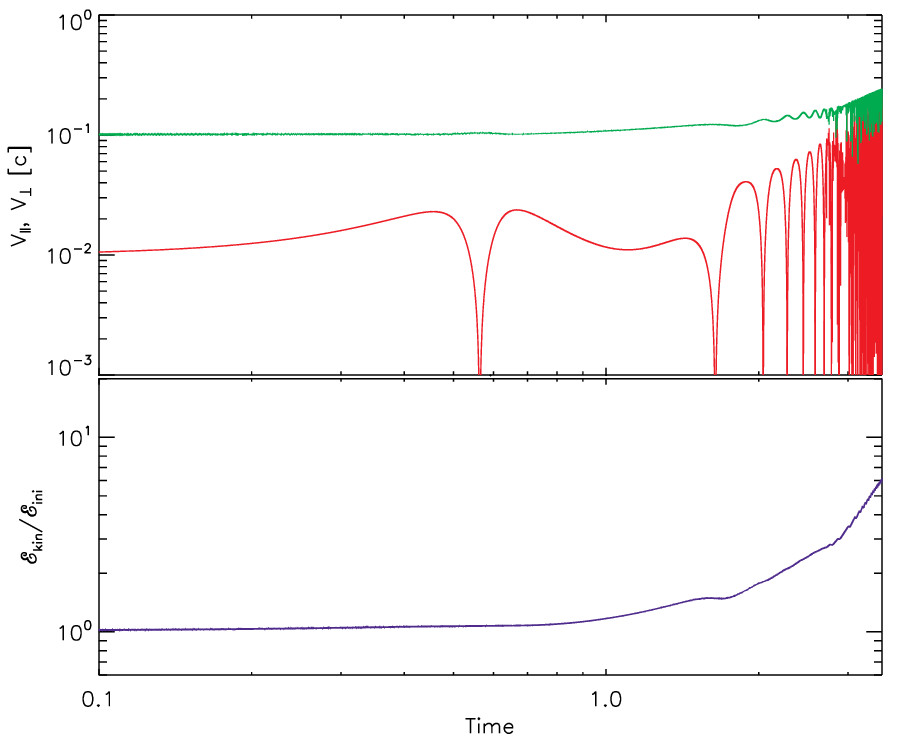}
 \caption{The case of acceleration near and within a single current sheet with a
Sweet-Parker configuration like
region (R6 in  Fig.~\ref{fig:locations}).  The left panel shows the trajectory of
a test proton approaching the diffusion region.  The color of the trajectory
corresponds to the particle energy (which increases from red to yellow and then
finally to white when the particle reaches the current sheet). The right panel
shows the evolution  of the particle energy.    In the model of
Sweet-Parker reconnection presented in this figure we used explicit large resistivity
coefficient $\eta = 10^{-3}$ in order to make reconnection fast.  The grid size in the model was set to $\Delta x =
1/1024$. (Extracted from \cite{kowal11}.)
\label{fig:sp_acceleration} }
\end{figure*}

Figure~\ref{fig:sp_acceleration} zooms in
 the details of the acceleration of a test particle near and within
a single (Sweet-Parker shaped; see also Section 3.4) current sheet.
Before the particle reaches the current sheet discontinuity it is drifted by the
plasma inflow and the increasing gradient of B as it approaches the current
sheet.  When it enters the discontinuity (the white part of the trajectory in
the left panel), it bounces back and forth several times and gains energy (which
increases exponentially as shown in the right panel of
Figure~\ref{fig:sp_acceleration}) due to head-on collisions with the converging
flow, on both sides of the magnetic discontinuity, in a first order
Fermi process, as described in GL05 {degouveia05}.  At the same time it drifts
along the magnetic lines which eventually allow it to escape from the
acceleration region.  Therefore, we see two mechanisms: a drift acceleration
(dominating outside of the current sheet) and first order Fermi acceleration
inside the current sheet.  These processes naturally depend on the initial
particle gyroradius, since it determines the amount of time the particle remains
in the acceleration zone before escaping.


\subsection{2D versus 3D simulations}
\label{sec:2d_vs_3d}

The results presented in the previous sections were obtained for 2D models
without a guide field.  This means that in this case the magnetic lines creating
the islands are closed and a charged particle can be trapped indefinitely in
such an island.  The presence of a guide field normal to the plane of
Figure~\ref{fig:locations} opens the magnetic loops and allows the charged
particles to travel freely in the out-of-plane direction.  Moreover, the islands
evolve much slower in the presence of a strong guide field.

\begin{figure*}[ht]
\center
\includegraphics[width=0.5\textwidth]{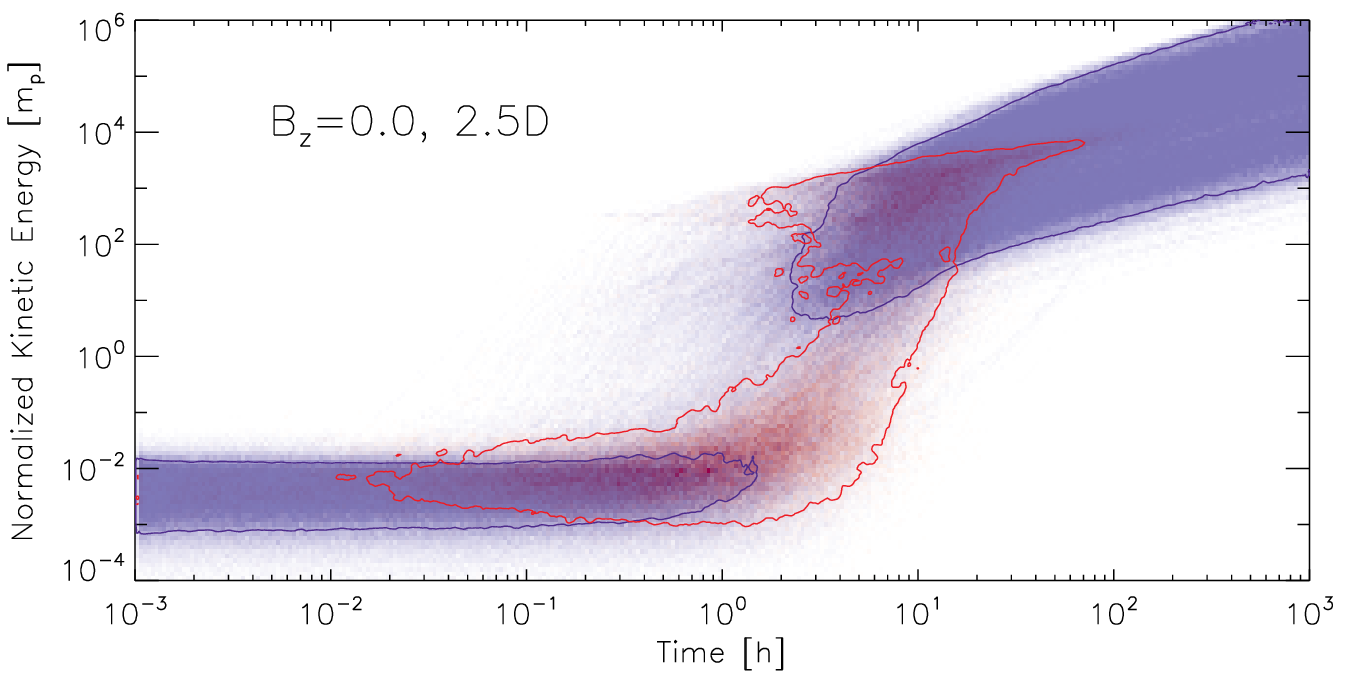}
\includegraphics[width=0.5\textwidth]{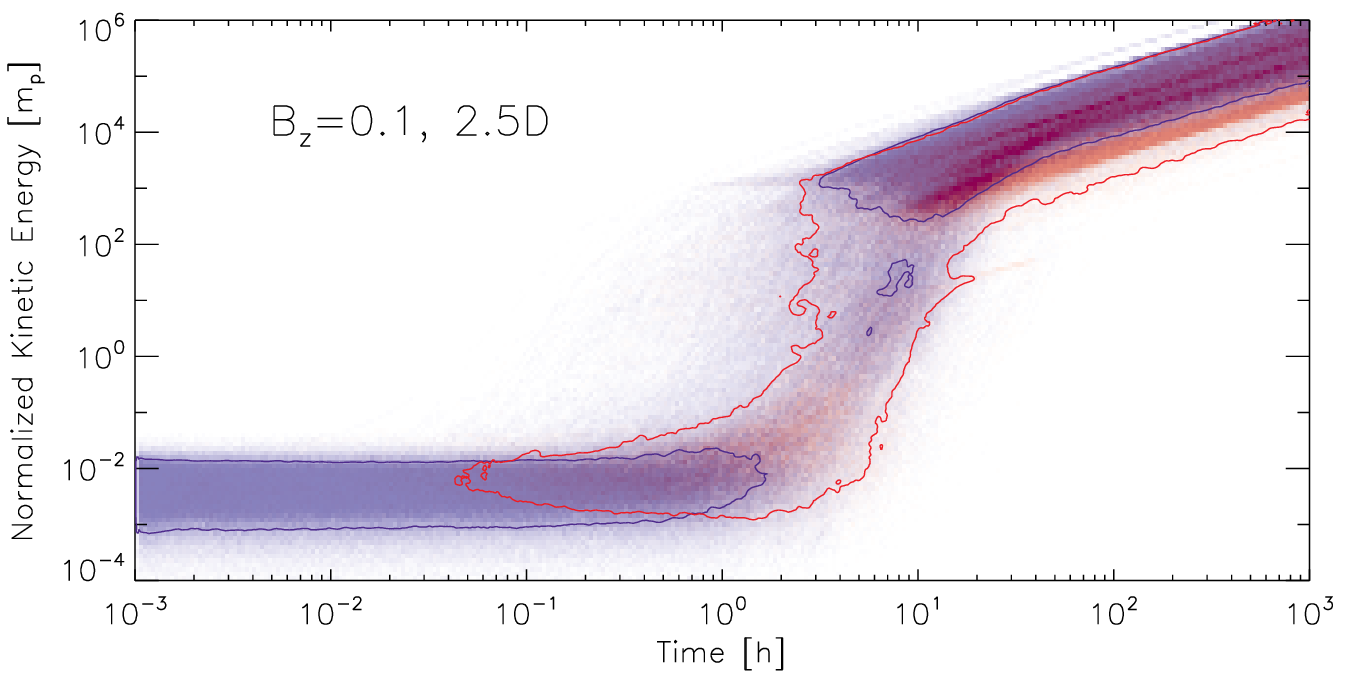}
\includegraphics[width=0.5\textwidth]{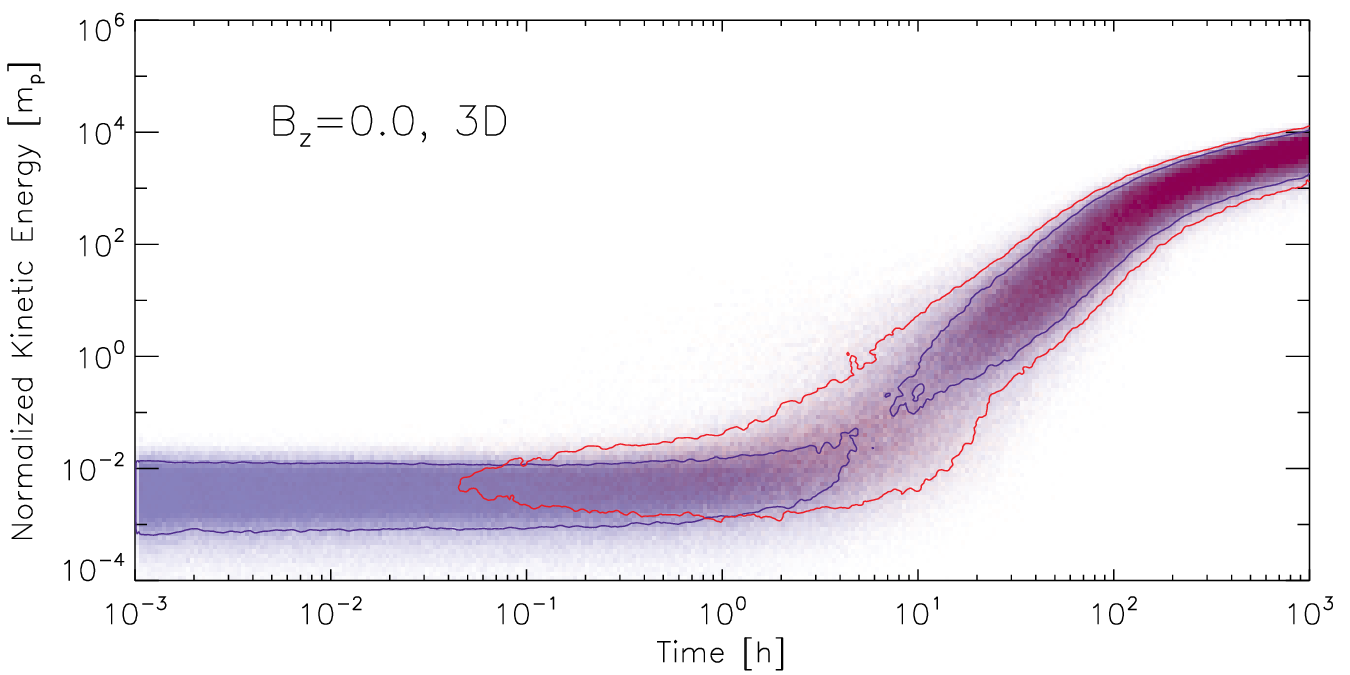}
\caption{Kinetic energy evolution of a group of 10$^4$ protons in 2D models of
reconnection with a guide field strength $B_z$=0.0 and 0.1 (top and middle
panels, respectively).  In the bottom panel a fully 3D model with initial
$B_z$=0.0 is presented.  The colors show how the parallel (red) and
perpendicular (blue) components of the particle velocities increase with time.
The contours correspond to values 0.1 and 0.6 of the maximum number of particles
for the parallel and perpendicular accelerations, respectively.  The energy is
normalized by the rest proton mass energy.  The background magnetized flow with
multiple current sheet layers is at time 4.0 in Alfv\'en time units ($t = L/V_A$, where L is the size of the computational domain and $V_A$ is the Alfv\'en speed corresponding to the initial magnetic field in the system) for  all
models. (From \cite{kowal11}.)
\label{fig:energy_2d_3d}   }
\end{figure*}

Figure~\ref{fig:energy_2d_3d} depicts the time evolution of the kinetic
energy of the particles which have their parallel and perpendicular (red and
blue points, respectively) velocity components accelerated for three models of
reconnection.
The kinetic energy
is normalized by the proton rest mass value, i.e., it is actually  $\gamma - 1)$
that is plotted, where $\gamma$ is the Lorentz factor.
 In the  2D model without
a guide field (as in the models studied in the previous section) there is an exponential growth of energy mostly
due to the acceleration of the parallel component which stops after the energy
reaches values of 10$^3$--10$^4$.
  From that level on, particles accelerate their perpendicular component
only with smaller linear rate in a log-log diagram.  In the 2D model with a weak guide field $B_z$=0.1 normal
to the plane of Figure~\ref{fig:locations}, there is also an
exponential acceleration of the parallel velocity component, but  due to the
presence of the weak guide field, this component accelerates further to
higher energies at a similar rate as the perpendicular one. This implies that
the presence of a guide field removes the restriction seen in the 2D model
without a guide field and allows the particles to increase their parallel
velocity components as they travel along the guide field, in open loops rather
than in confined 2D islands. This result is reassured by the 3D model in  Figure~\ref{fig:energy_2d_3d}, where no guide field is necessary
as the MHD domain is fully three-dimensional. In this case, we clearly see a
continuous increase of both components, which suggests that the particle
acceleration behavior changes significantly when 3D effects are considered, i.e.
where open loops replace the closed 2D reconnecting islands.

With the   parametrization considered, the gyroradius of
a proton becomes comparable to the size of the box domain when its Lorentz factor
reaches a value of a few times 10$^4$.  The largest islands in the system can
have sizes of a few tenths of the size of the box.  These rough estimates help
us to understand the energy evolution in Figure~\ref{fig:energy_2d_3d}.   In the case with no guide field (top panel of
Fig.~\ref{fig:energy_2d_3d}), the exponential parallel acceleration stops right
before the energy value 10$^4$ is reached.  After this, the rate of acceleration
significantly decreases.  This occurs because the Larmor radius of the particles
has become larger than the sizes of biggest islands.  Therefore, from this level
on the particles cannot be confined anymore within the islands and the first
order Fermi acceleration ceases. 
After that, there is a much slower drift
acceleration (of the perpendicular component only) caused by the gradients of
the large scale magnetic fields and acceleration between islands.  If a guide field is inserted in such a system (as in the model of the  middle panel  of
Fig.~\ref{fig:energy_2d_3d}),
the picture is very similar.  However, since the particles
are now able to travel along the guide field, their parallel velocity component also
continues to increase after the 10$^4$ threshold.  Of course, in the 3D model, the particles follow
the same trend (bottom panel of Fig.~\ref{fig:energy_2d_3d}).

While in two dimensional MHD models without a  guide field
 the parallel acceleration saturates at some level,  in the presence of an out-of-plane guide field or in
three dimensional models this saturation effect is  removed.

\subsection{Acceleration in 3D Sweet-Parker Reconnection}
\label{sec:sweet-parker-reconnection}

In the Sweet-Parker model of  reconnection of two large scale magnetic fluxes of opposite polarity  \cite{sweet58,parker57}, the speed of reconnection, i.e. the speed at which  two inflowing  magnetic field  lines annihilate by ohmic dissipation, is roughly $\eta/\Delta$, where $\Delta$ is the width of the current sheet discontinuity (Figure 1) and $\eta$ is the Ohmic magnetic resistivity.  The entrained plasma follows the local field lines and exits through the edges of the current sheet at roughly the Alfv\'en speed, $V_A\equiv B/(4\pi \rho)^{1/2}$, where $\rho$ is the local density. Thus using momentum flux conservation it is easy to demonstrate that the resulting reconnection speed is a tiny fraction of the Alfv\'en speed,
 or  $V_{R}\approx V_A S^{-1/2}$, where $S = L V_A / \eta$ is the
Lundquist number and   $L$ is the length of the current sheet.  Due to the typically huge astrophysical sizes of the reconnection sites, $S$ is
also huge for Ohmic diffusivity values (e.g., for the interstellar medium, $S \sim10^{16}$) and this makes
the Sweet-Parker reconnection  very slow.
 However,  observations require a reconnection speed close to $V_A$ in several circumstances (e.g., in solar flares). A way to speed up reconnection is to invoke plasma instabilities, as for instance, the stream instability which makes resistivity anomalously large in the relation above \cite{parker79}. Another way is to consider the presence of turbulence in the current sheet \cite{lazarian99}, a process that  will be described  in \ref{sec:stochastic-reconnection}.

\begin{figure*}
\center
\includegraphics[width=0.45\textwidth]{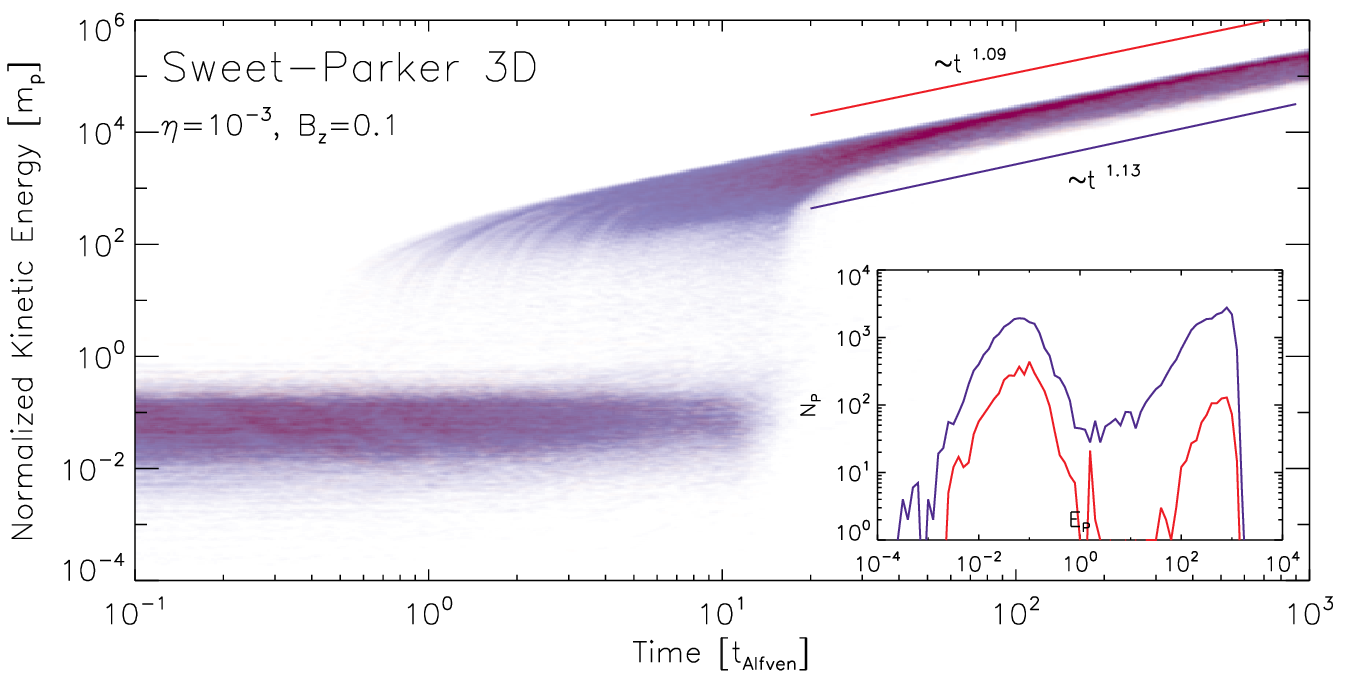}
\includegraphics[width=0.45\textwidth]{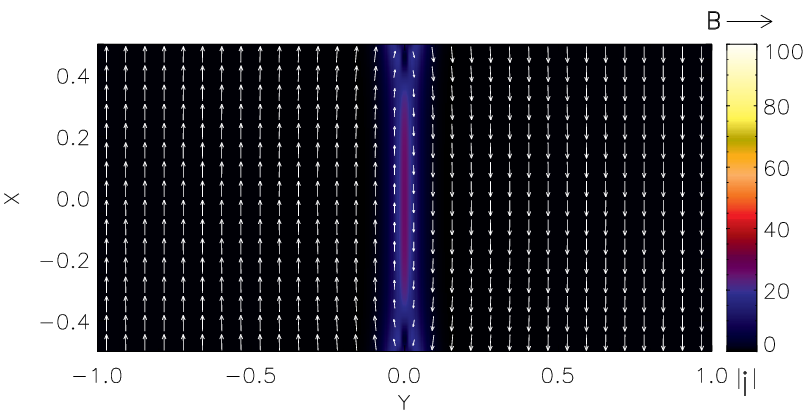} \\
\includegraphics[width=0.45\textwidth]{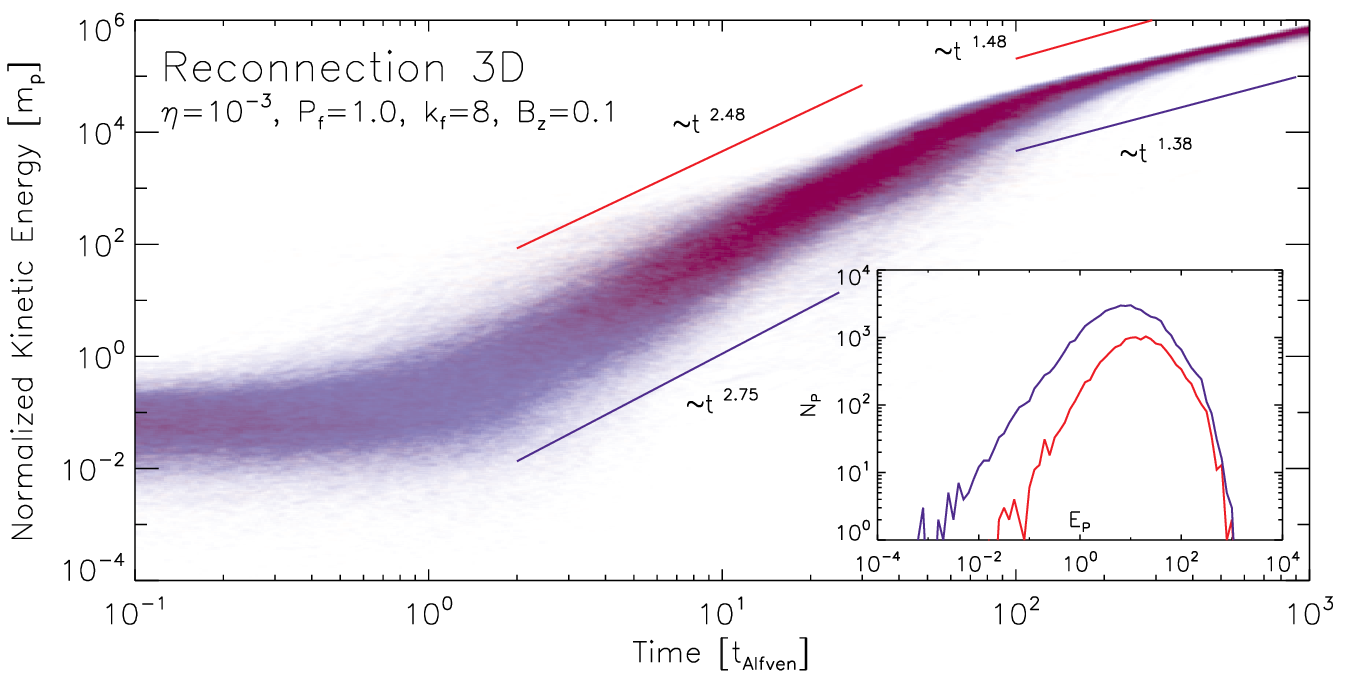}
\includegraphics[width=0.45\textwidth]{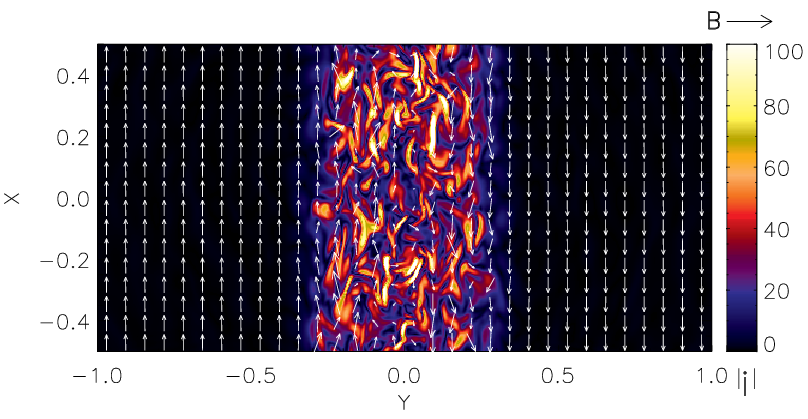} \\
\includegraphics[width=0.45\textwidth]{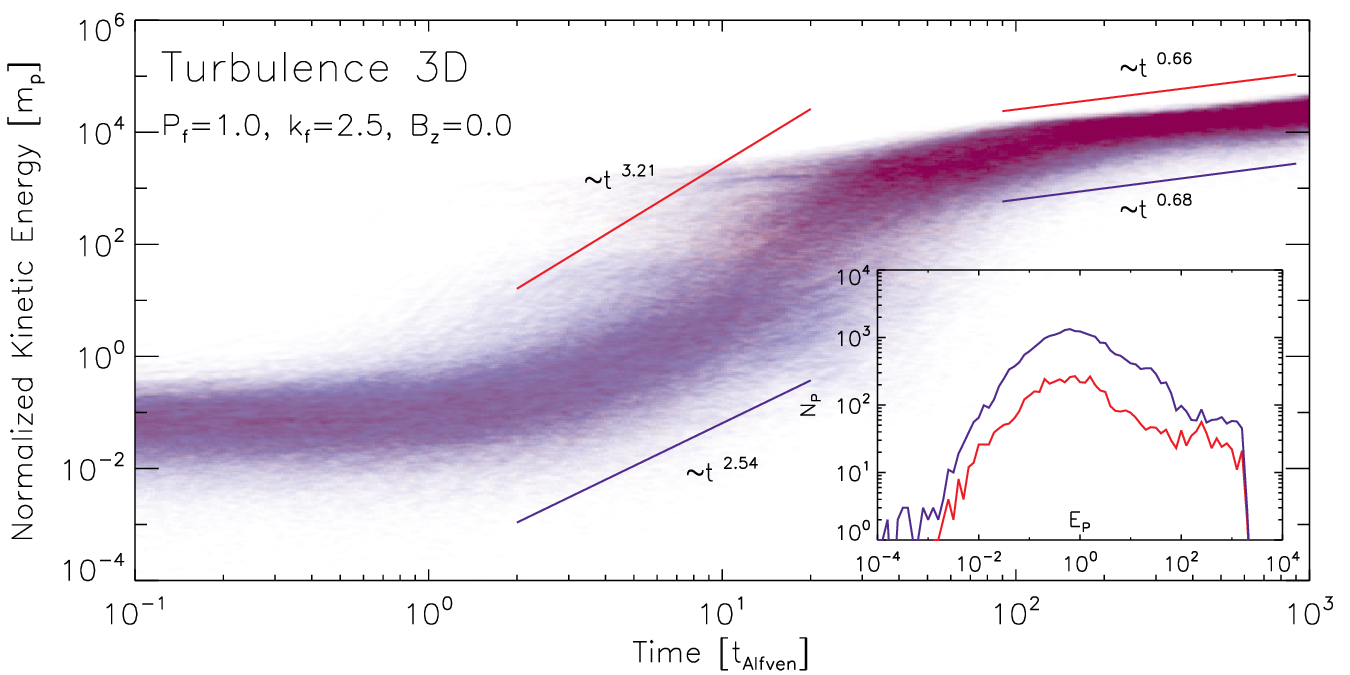}
\includegraphics[width=0.45\textwidth]{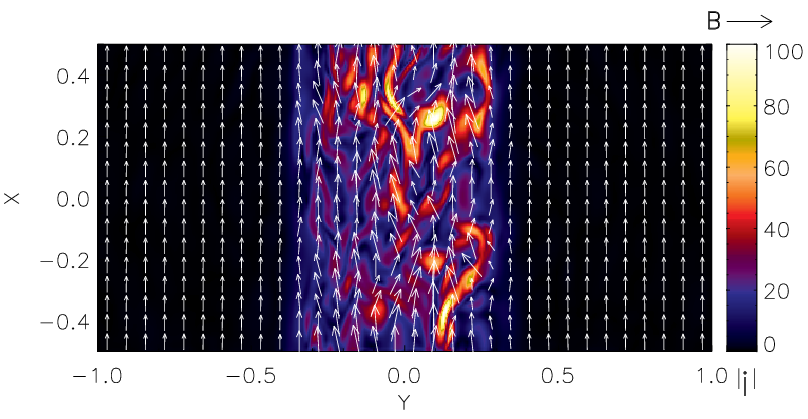}
\caption{{\em Left column:} Particle kinetic energy distributions for 10,000
protons injected in the Sweet-Parker reconnection (top), fast magnetic
reconnection (middle), and purely turbulent (bottom) domains.  The colors
indicate which velocity component is accelerated (red or blue for parallel or
perpendicular, respectively).  The energy is normalized by the rest proton energy.
Subplots show the particle energy distributions at $t=5.0$.  {\em Right column:}
 XY cuts through the domain at $Z=0$ of the absolute value of
current density $|\vec{J}|$ overlapped with the magnetic vectors for the
Sweet-Parker reconnection (top), fast reconnection (middle), and purely
turbulent domains (bottom).  For the top and middle models with large scale current sheets it was employed $B_{0z} = 0.1$, $\eta=10^{-3}$, and a
resolution 256x512x256, while for the bottom pure turbulent model  it was employed  $B_{0z} = 0.2$ and a resolution
128x256x128. (From \cite{kowal12}.)
\label{fig:energy_evolution}}
\end{figure*}

In the model shown in the top  of
Figure~\ref{fig:energy_evolution}, KGL12 \cite{kowal12} investigated the acceleration of thousands of particles in a Sweet-Parker current sheet but, in order to make reconnection fast, they  employed a diffusivity coefficient $\eta =
10^{-3}$ expressed in code units which,  due to the numerical diffusivity,
is several orders of magnitude larger than the typical Ohmic diffusivity in
astrophysical environments and besides, makes the Sweet-Parker reconnection in the
simulation efficient.  The time evolution of the energy distribution for the
accelerating particles is shown for this model in the top left panel of
Figure~\ref{fig:energy_evolution}.  Initially, the perpendicular acceleration
dominates, because the volume in which particles are injected is much larger than
the current sheet.  The
perpendicular acceleration, due to a drift of the magnetic flux, starts before
the particles reach the reconnection region \cite{kowal11}.  The distribution of
particles does not change significantly until $t=1.0$.  Then, a rapid increase
in energy by roughly four orders of magnitude appears for a fraction of
particles.  We observe a big gap between the energy levels before and after
these acceleration events, which is also evident in the particle energy spectrum
depicted in the  subplot of the same diagram.  The events are spread in
time because particles gain substantial energy at different instants  when
crossing the current sheet.  The energy growth during this stage is exponential.
This is clearly due to the first-order Fermi acceleration process, as stressed before
\cite{degouveia05, kowal11, kowal12} and already shown in Figure~\ref{fig:sp_acceleration}.
We note that, as in Figure \ref{fig:energy_2d_3d}, the particles accelerate at smaller rates after reaching the energy level $\sim 10^4$, because the thickness of the acceleration region becomes smaller than their Larmor radii and this is consistent with the predictions of eq. 11.  

Although the Sweet-Parker model with an artificially enhanced resistivity
results in a predominantly first-order Fermi acceleration, only a small fraction
of the injected particles is trapped and efficiently accelerated in the current
sheet (see the energy spectrum of the accelerated particles in the bottom right
of Figure~\ref{fig:energy_evolution}). This is because the acceleration zone is very
thin.

\subsection{Acceleration in 3D reconnection sites with Turbulence}
\label{sec:stochastic-reconnection}

As remarked before, Lazarian \& Vishniac \cite{lazarian99} (see also the \textbf{Chapter by Lazarian} in this book) proposed a model for fast reconnection
that does not depend on the magnetic diffusivity (see also \cite{eyink11}).  Given the fact that MHD turbulence is ubiquitous in  astrophysical environments,  this may  be  a universal trigger
of fast reconnection.  The predictions of this model have been tested successfully
 by numerical simulations \cite{kowal09,kowal12} which confirmed that the
reconnection speed is of the order of the Alfv\'en speed and independent of resistivity.  An important
consequence of  fast reconnection by turbulent magnetic fields is the
formation of a thick volume filled with small scale magnetic fluctuations.  In
order to test the acceleration of particles within such a domain, KGL12 \cite{kowal12} introduced
turbulence within a current sheet with a Sweet-Parker configuration (as
described in the previous paragraph) and followed the trajectories of
10,000 protons injected in this domain.

The middle left panel of Figure~\ref{fig:energy_evolution} shows the evolution
of the kinetic energy of the particles in this case.  After injection, a large fraction of
test particles accelerates and the particle energy growth occurs earlier than in
the Sweet-Parker case (see also the energy spectrum at $t=5$ in the detail at
the bottom right of the same diagram).  This is explained by a combination of
two effects: the presence of a large number of converging small scale current
sheets and the broadening of the acceleration region due to the turbulence.
Here, we do not observe the gap seen in the Sweet-Parker reconnection, because
particles are continually accelerated by encounters with several small and
intermediate scale current sheets randomly distributed in the thick volume.  The
acceleration process is clearly still a first order Fermi process, as in the
Sweet-Parker case, but more efficient as it involves larger number of particles,
since the size of the acceleration zone and the number of scatterers have been
naturally increased by the presence of turbulence.

An inspection of the particle spectrum in the subplot of the middle panel of  Figure~\ref{fig:energy_evolution} at t=5 c.u.  reveals already the formation of 
a power law spectrum $N(E) \sim E^{-1}$  in the energy range $E/m_pc^2 \sim 10-10^3$, where $m_p$ is the proton mass. 
This power law index is compatible with former results obtained from collisionless  PIC simulations (e.g., \cite{zenitani01}).

\subsection{Acceleration by Reconnection in Pure 3D Turbulent Environments}
\label{sec:turbulence}

The bottom left panel of Figure~\ref{fig:energy_evolution}
shows the kinetic energy evolution of accelerated particles in a domain with
turbulence only, i.e., without large scale magnetic flux tubes and thus no large scale  current sheet.  This could be the situation in typical diffuse MHD environments like the interstellar, the intracluster and intergalactic media. One of the fundamental points of the
 Lazarian \& Vishniac  theory \cite{lazarian99} is the fact that whenever there is MHD turbulence, there will be fast  reconnection of the turbulent magnetic field lines from the injection to the dissipation scales of the turbulence \cite{lazarian11, lima10, lima12, lima13, leao13}. Therefore, particles will be able to accelerate while trapped within these multiple current sheets at all scales.

We see in Figure~\ref{fig:energy_evolution}  that the acceleration is less efficient at the beginning and a much smaller fraction
of particles is accelerated than when a large scale current sheet is present, as in the middle panel of Figure~\ref{fig:energy_evolution}.  In the later case, the converging flow on both sides of the large scale current sheet
 brings  approaching scattering centres that undergo only head-on collisions with the particles allowing a continuous growth of the particle
energy until the saturation level.  In pure turbulence, however, the absence of a large scale
converging flow results in a random particle scattering into both approaching and
receding small scale magnetic fluctuations (although at a smaller rate), so that the
overall acceleration is possibly a second-order Fermi process.

It should be also remarked that earlier studies of particle acceleration in pure 3D MHD turbulent environments   have already identified the development of stochastic acceleration and a power-law tail in the energy spectrum of the accelerated particles (see e.g., \cite{dmitruk03}).


\section{Discussion and Conclusions}

Advances both in the understanding of magnetic reconnection in the MHD regime and improvement on high energy observations
have lately motivated the studies of particle acceleration in reconnection sites of astrophysical sources and environments.

In this Chapter, we reviewed   particle acceleration in 2D and 3D (collisional) MHD domains of
magnetic reconnection.  It has been shown  that particles can be efficiently accelerated by reconnection through a first order Fermi process within large scale current sheets (built up  by large scale converging magnetic fluxes), specially when local turbulence is present. The later makes the reconnection fast \cite{lazarian99} and   the volume of the accelerating zone thick  \cite{kowal11, kowal12}. The particles trapped within the current sheet   suffer
several head-on scatterings with the contracting  magnetic fluctuations as originally predicted by GL05 \cite{degouveia05} (see also \cite{drake06}) and undergo an exponential growth in their kinetic energy, as demonstrated numerically by KGL11  \cite{kowal11} and KGL12 \cite{kowal12}.
 In a Sweet-Parker configuration (with the reconnection speed made
artificially large by numerical diffusion) the acceleration rate
is slightly smaller because of the thinner current sheet, but   it is also a
first-order Fermi process.
In contrast, in  pure 3D turbulent environments (with no large scale current sheets),  particles with gyroradii
smaller than the injection scale of the turbulence are accelerated  through a second order Fermi
process while interacting with both approaching and receding small scale turbulent current sheets. This process can be particularly important for cosmic ray acceleration in diffuse turbulent environments like the interstellar, intracluster and intergalctic media, while the first order Fermi acceleration in large scale current sheets can be relevant particularly in stellar coronae, compact sources (like the accretion disk coronae in AGNs, microquasars, etc.; see Figure 7 as an illustrative example), and  highly magnetized flows (like AGN, microquasar and GRB jets).

\begin{figure*}
\center
\includegraphics[width=1.2\textwidth]{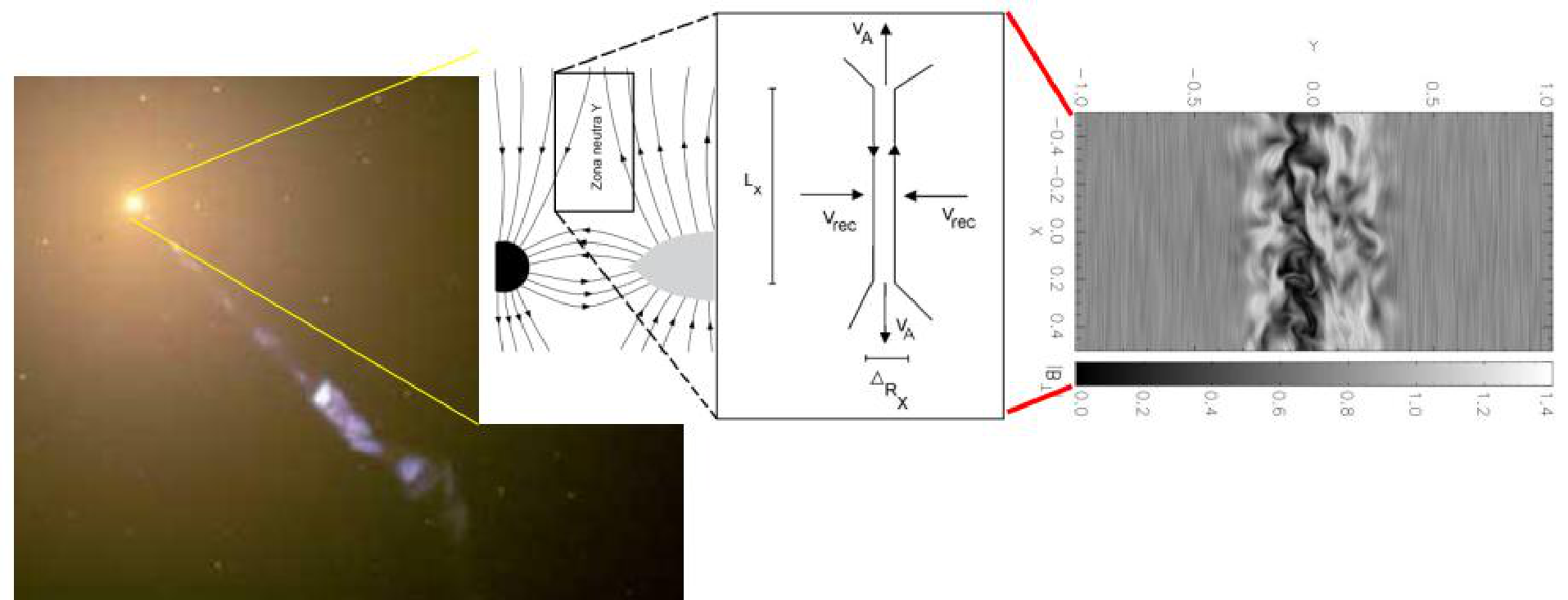}
\caption{
From left to right the figure shows: the HST image of M87 AGN; a schematic representation of the expected magnetic field structure around the accretion disk and the central black hole (as in \cite{degouveia05}); a schematic representation of the reconnection zone with the two converging magnetic fluxes of opposite polarity as in a Sweet-Parker configuration approaching each other with a reconnection speed $V_{R} = V_{rec}$, and a 3D MHD simulation of magnetic reconnection with turbulence injected within the current sheet to make reconnection fast (as in \cite{kowal12}).
\label{fig:schem}}
\end{figure*}

It  has been  also shown  that the acceleration within fast reconnection sites
 works both, in collisionless  and collisional (especially in the presence of turbulence)
 environments.  The acceleration by magnetic reconnection in the 2D collisional MHD
regime \cite{kowal11, kowal12} successfully reproduces the results
obtained  with  (collisionless) 2D PIC codes (e.g. \cite{drake10, drake12}). This proved that   the
acceleration in reconnection regions is a universal process which is not
determined by  details of the plasma physics or kinetic effects.
However, in the collisional case,  only the injected
particles with Larmor radii near the MHD scales are effectively accelerated.
This injection problem can be solved using hybrid codes able to  resolve both
the kinetic and the MHD scales.

It should be noticed  that  \cite{onofri06} also investigated particle acceleration
in MHD reconnection regimes. However, they concluded that MHD would not be a good approximation
to describe the process of acceleration by reconnection.  This is because their 3D numerical
simulations were performed in a fully resistive MHD regime. Therefore,  they
obtained an efficient particle acceleration due to the high electric field
induced by the resistivity term only (see Eq. 11) and an absorption of most of the available magnetic
energy by the electrons in a very small fraction of the characteristic time of
the MHD simulation.  This led them to conclude that resistive MHD codes are
unable to represent the full extent of particle acceleration in 3D reconnection.
 KGL11 and KGL12 \cite{kowal11, kowal12} on the other hand,  explored  particle acceleration in a nearly
ideal MHD regime where only small numerical resistivity was present.  In this case, the
contribution of a resistivity induced electric field is negligible when compared
to the advection component, namely, the electric field resulting from the plasma
motion in the magnetized medium, $- \vc{v} \times \vc{B}$.

It has been also demonstrated that the acceleration of energetic particles in 2D and 3D reconnection domains shows substantial
differences \cite{kowal11}. This calls for focusing on realistic 3D geometries of reconnection.
The numerical studies \cite{kowal11, kowal12} have also revealed  that apart from the first order Fermi acceleration,
additional less efficient acceleration processes, like drift acceleration due to non-uniform magnetic fields and second order Fermi, also interfere in the process (see also \cite{dmitruk03}).

All these numerical studies  of particle acceleration  have neglected  the time evolution of the MHD environment. This is in general valid since this is much longer than the particle time scales. In fact, when considering, for instance, the acceleration within large scale current sheets with turbulence,  particles are accelerated by magnetic fluctuations in the turbulent field and interact resonantly with larger and larger structures as their energy increases due to the scatterings. In a steady state turbulent environment, as considered here, particles will see on average the same sort of fluctuation distribution, so that after several Alfv\'en times, one should expect no significant changes in the particle spectrum due to the evolution of the large scale MHD environment \cite{delvalle13}. Nonetheless, this evolution may be important when considering more realistic non-steady environments and when calculating real spectra and loss effects (e.g. \cite{lehe09, khiali13}).
It may be also relevant when considering the (second order Fermi) acceleration in pure turbulent environments (as in Figure~\ref{fig:energy_evolution}, bottom panel). In this case,  electric fields arising from slow modes (betatron
acceleration) can be relevant  to the acceleration process making it twice as larger
 since the betatron term
$\partial\vc{B}/\partial t$ contributes as much as the electric field term (Eq. 11) in the second
order process.  In  forthcoming studies when considering more realistic
non-steady environments,  MHD data cubes varying in time should be used specially in pure turbulent studies.

It should be remarked also that the collisional MHD  simulations shown here focussed on proton acceleration. Although  applicable to electrons too,  the numerical integration of the electron trajectories is much longer. Nevertheless, such tests are also needed.
In particular, it has  been  suggested that  in
 electron-positron plasmas  the pairs could  annihilate in
compressed reconnection sheets \cite{drury12}, so that this could  influence the acceleration by reconnection, e.g.,  in pulsar winds and relativistic jets in general.

\subsection*{Analytical predictions versus numerical simulations }
It has been seen that analytical studies of the first order Fermi process in large scale current sheets   predict that \cite{degouveia05, drury12}: (i) the acceleration rate is similar to that for shock acceleration; and (ii) the energy power law spectrum of the accelerated particles can be even harder \cite{drury12} than the one predicted for shock acceleration
 and independent on the reconnection velocity  \cite{degouveia05}.
 These predictions, although based on very simplified assumptions can be, in principle,  tested with   numerical simulations. However,  a larger parametric space considering, e.g., different ratios between the initial Alfv\'en (or  reconnection) speed and the light speed, and different amplitudes of the injected turbulence  (which speeds up reconnection within the large scale current sheet) must be still performed  in order to assess the sensitivity of both the acceleration rate and the particle spectrum to the physical conditions in the reconnection domain.
 Results from collisional MHD numerical simulations with  injection Alfv\'en velocities in the range $v_A/c \sim 1/1000- 1/5$  
 \cite{kowal11, kowal12, delvalle13}, indicate that the acceleration time is nearly independent of the initial Alfv\'en (and   reconnection) speed
and is given by  $t_{acc} \sim E^{0.4}$ (this is directly derived  from diagrams as in Figures 5 and 6). This is initially longer than the estimated time by eq. 9,  but becomes comparable to it as the particles approach the maximum energy value that they can reach in the acceleration zone  \cite{delvalle13}. As we have seen, this maximum energy is attained when the particle Larmour radius becomes comparable to the size of the acceleration zone.


The determination of  the dependence  of the energy spectral index and the fraction of accelerated particles at high energies with  the initial and the boundary conditions is more complex (see Figure \ref{fig:energy_evolution}) and  requires further numerical studies, particularly considering the effects of particle feedback. So far, the results of  particle acceleration in 3D MHD reconnection sites indicate 
a hard power law spectrum $N(E) \sim E^{-1}$ (\cite{kowal11, kowal12}; see Section 3). This is  comparable with results obtained from 2D collisionless PIC simulations considering merging islands  (for which an energy power-law index $\sim (-1.5)$ has been found; \cite{drake12}), or  X-type Petschek's configurations   (e.g., \cite{zenitani01}, for which an energy power-law index  $\sim (-1)$ has been obtained).  






\begin{figure}[h]
\includegraphics[width=0.6\columnwidth]{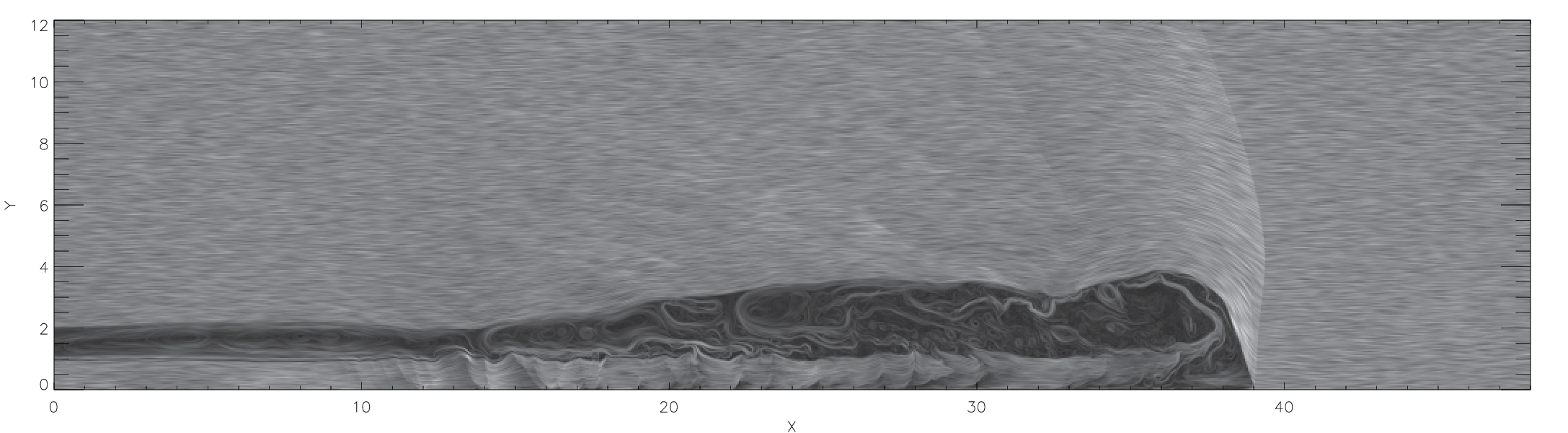} \\
\includegraphics[width=0.6\columnwidth]{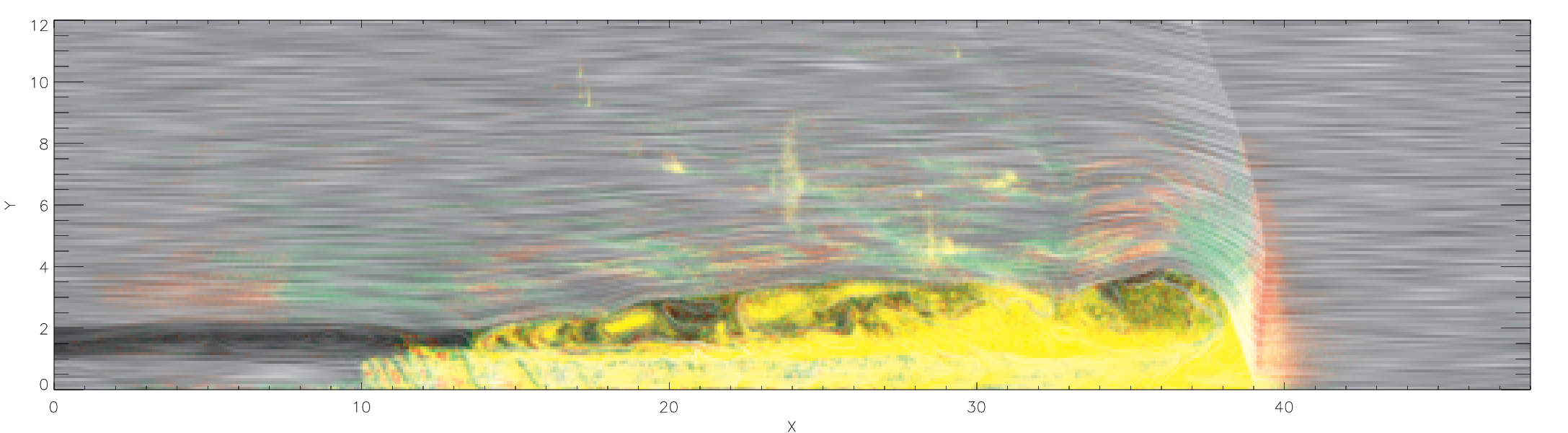}
\includegraphics[width=0.4\columnwidth]{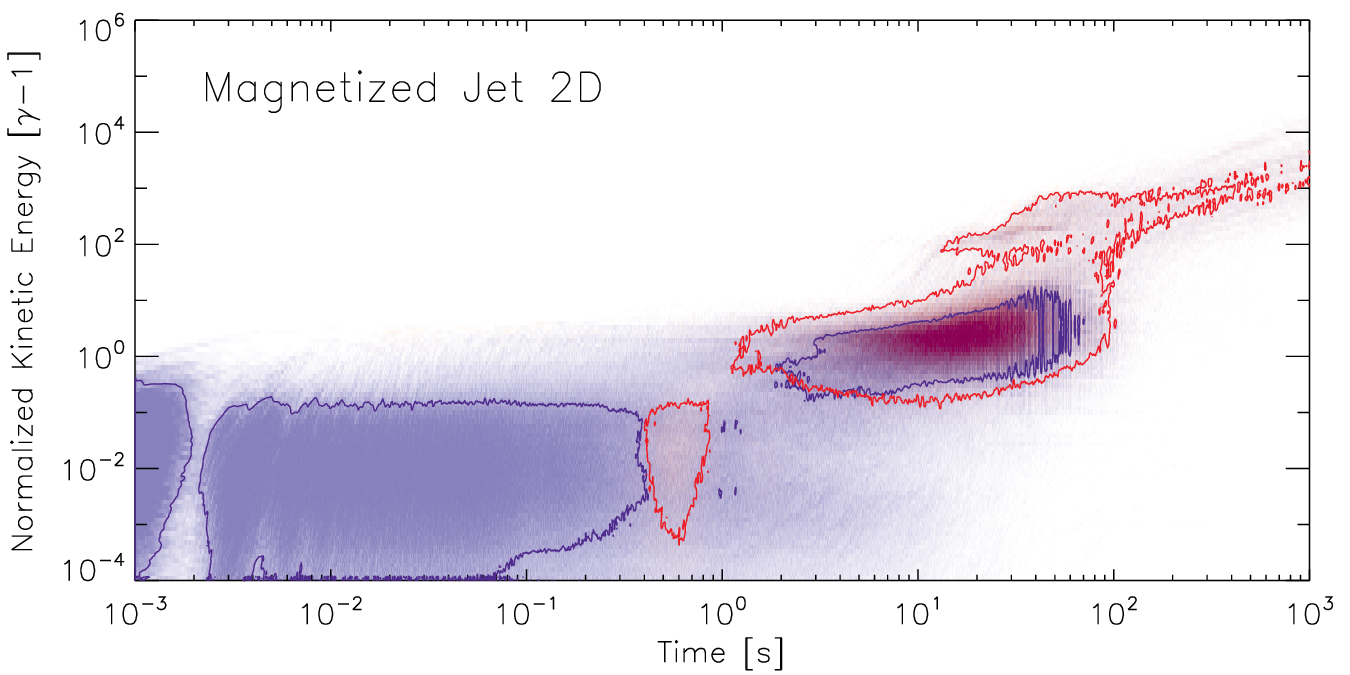} \\
\includegraphics[width=0.6\columnwidth]{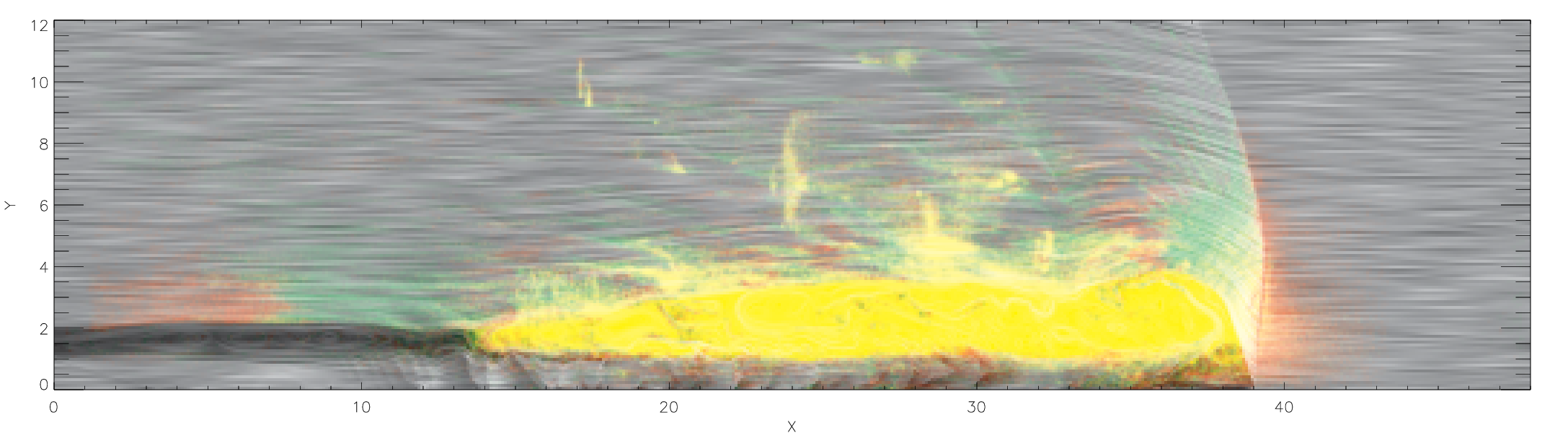}
\includegraphics[width=0.4\columnwidth]{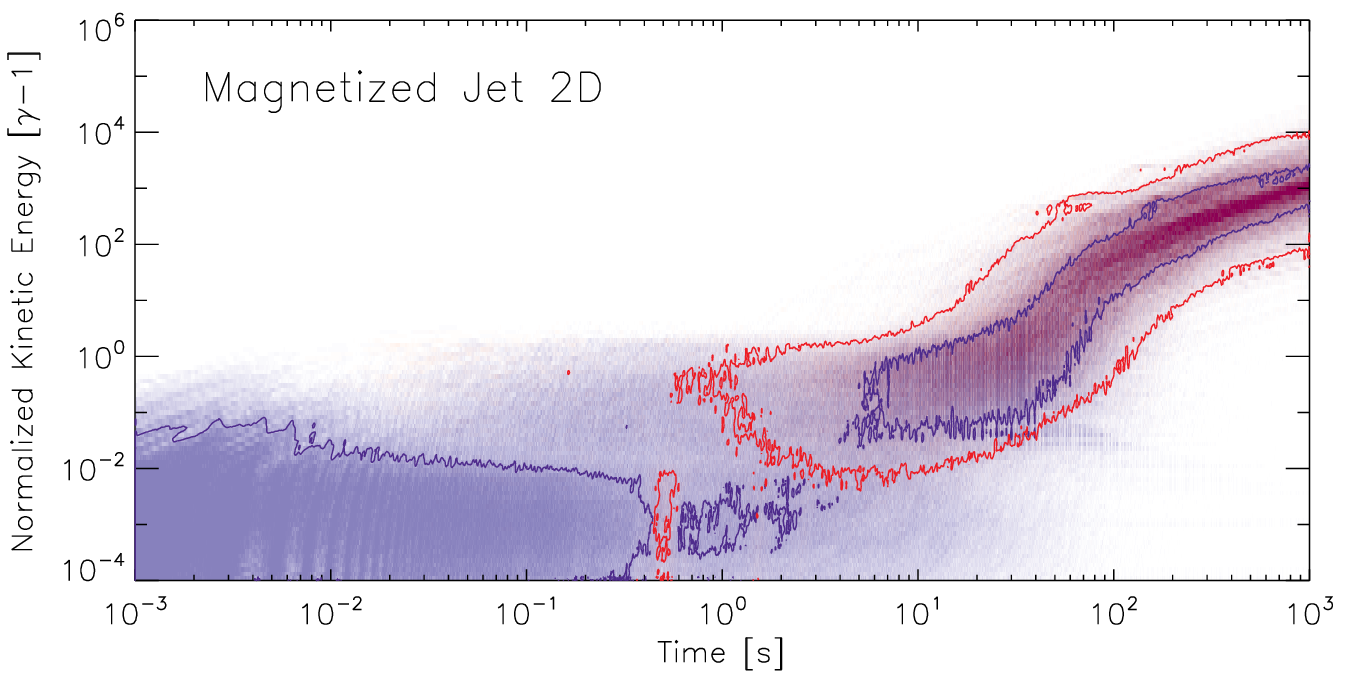}
\caption{2D relativistic-MHD jet. Left diagrams: top panel depicts the topology of the magnetic field represented as gray texture for a 2D relativistic jet at $t = 80.0$ in code units with initial uniform longitudinal magnetic field corresponding to a ratio between the gas pressure and the magnetic pressure $\beta = 1/300$, a density ratio between the jet and the ambient medium $0.01$; a jet Mach number $M_j=6.0$, and a Lorentz factor $\Gamma = 10.0$. The left middle and bottom panels show the same diagram but with superimposed semi-transparent color maps representing locations where the parallel and perpendicular particle velocity components are accelerated.  The red and green colors correspond to regions where either parallel or perpendicular acceleration occurs, respectively, while the yellow color shows locations where both types of acceleration occur.  The top left panel clearly depicts the large scale features typically seen in 2D jet simulations, i.e., the bow shock at the head where the light jet beam impacts supersonically a much denser environment, and internal shocks (or knots)
all along the beam which are driven by the pinch mode of the Kelvin-Helmholtz (K-H) instability.  We also identify a more turbulent cocoon enveloping the beam which is formed by the  mixing of the shocked ambient and jet gas. Magnetic islands can be clearly distinguished in this region. The left middle panel depicts only the accelerated particles within the beam and in the bow shock region, while the bottom left panel depicts the particles which are accelerated mostly in the surrounding cocoon. The right diagrams show the acceleration rates: the top one corresponds to the middle left panel, i.e., to the acceleration regions within the beam and the bow shock and therefore, is dominated by first order Fermi acceleration behind shocks; and the bottom right panel corresponds to the bottom left panel, i.e., to the acceleration mostly in magnetic islands and thus is dominated by first order Fermi due to magnetic reconnection.    The simulation was performed with a resolution 8192x4096.   10,000 test particles
were injected in this snapshot with an initial thermal distribution with a temperature corresponding to the sound speed of the relativistic-MHD model. (From \cite{kowal13b}; in preparation). \label{fig:RMHDjet}}
\end{figure}


\subsection*{Particle acceleration in relativistic domains of reconnection}
In this review,  we discussed mostly  Fermi  acceleration considering non-relativistic reconnection environments, that is, generally assuming $V_{R}$   smaller than the light speed. This seems to be  appropriate in the solar (or stellar) corona and wind and in the Earth magnetotail where this mechanism has been more extensively explored. However, in systems like, e.g, the very near surrounds of black holes and pulsars, $v_A \sim  c$  and thus,  since  in fast reconnection  $V_R$ must  approach  $v_A$, reconnection itself may become relativistic in such domains. 
There has been some advancement in relativistic reconnection studies too. The 
theoretical grounds have been established by a number of authors (\cite{lovelace92, blackman94, lyubarsky05, lyutikov03, jaroschek04, hesse07, zenitani08, zenitani09, komissarov07, coroniti90, lyubarsky01}; see also \cite{uzdensky11a} and references therein for a review and the chapter by Lazarian et al. in this volume).  
Essentially, it has been realized that  Sweet-Parker reconnection in the relativistic regime is slow, as in the non-relativistic regime,  while  2D X-point Petschek's  reconnection  predicts a fast  rate as in the non-relativistic domain \cite{lyubarsky05}.  The numerical advances  in relativistic reconnection have been performed so far only for 2D collisionless X-point  Petschek’s  configurations  by means of PIC simulations of pair plasmas, but have confirmed the results of the  analytical theory.  In such relativistic collisionless electron-positron pair plasmas, the investigation of relativistic particle acceleration  is almost  straightforward. Studies by, e.g.,  Zenitani \& Hoshino (2001) \cite{zenitani01} have revealed results which are compatible with those of acceleration in non-relativistic reconnection.
 
However, studies of particle acceleration in the collisional relativistic MHD  regime (RMHD) are still in their childhood. As an example,  Figure \ref{fig:RMHDjet} shows very preliminary results of simulations of 10,000  test particles injected in a 2D relatvistic MHD jet system \cite{kowal13b}.  The colors highlight the regions where particles are being accelerated, mostly through  first order Fermi, to relativistic energies both behind shocks and within  magnetic reconnection islands.  The comparison between  the acceleration rate diagrams on the right hand side of this figure  suggests that both mechanisms are competitive \cite{kowal13b}. 

These results are encouraging and may have rather important consequences  on particle acceleration and high energy emission processes  in microquasars, pulsar winds, AGNs, and  GRBs  and demand further extensive investigation. 
The  assessment  of the role  of this mechanism   in modelling flares in the spectrum
 of  compact sources is also in order \cite{degouveia05, degouveia10a, degouveia10b, drury12}.
For instance, in recent work Cerutti et al. (2013) \cite{cerutti13}  performed 2D PIC collisionless reconnection simulations of ultra-relativistic pair plasmas considering the effects of 
radiation reaction on the particles. They detected several features of  first order Fermi particle  acceleration in their simulations that had been already revealed in non-relativistic simulations (e.g. \cite{kowal11, kowal12}), but  were also able to reproduce the observed spectral energy distribution of the Crab nebula flares. 
Another  recent  study  have demonstrated that the acceleration by magnetic reconnection   can be more efficient than shock acceleration in the surrounds of galactic black holes (microquasars) and reproduce  the observed high energy spectral distribution of sources like Cyg X-1 and Cyg  X-3 \cite{khiali13}. In fact, acceleration by magnetic reconnection seems to have  a major role in  the surrounds of  a broad range of black hole sources  (from microquasars to low luminous AGNs) \cite{degouveia10a,kadowaki14}.

Finally,   particles accelerated in  domains of reconnection, particularly in pure turbulent regions,
may be available as  seed populations for further first order Fermi shock or magnetic reconnection acceleration  in  these different  systems.



To summarize,
 magnetic reconnection is now recognized as an essential process not only in the solar system but also beyond it, in a large number of astrophysical sources, including   turbulent environments which in turn, are  ubiquitous.  In this situation the acceleration of
particles by reconnection may play a vital role, the   importance of
which should be evaluated with further extensive research.

\begin{acknowledgement}
 E.M.G.D.P. acknowledges  partial support from the Brazilian agencies FAPESP (grant no. 2006/50654-3 and CNPq  (grant no. 300083/94-7) and G.K. also acknowledges support from FAPESP (grant 2009/50053-8). The authors  are also in debt to Alex Lazarian for fruitful and joyful collaboration in this work. Part of the simulations presented here have been carried out in LAI (Astrophysical Laboratory of Informatics at IAG-USP).
\end{acknowledgement}



\end{document}